\documentclass[preprint,12pt,authoryear]{elsarticle}

\usepackage{amssymb}
\usepackage{url}
\usepackage{multirow}
\usepackage{physics}

\usepackage{array}
\renewcommand{\arraystretch}{1.1}
\newcolumntype{C}{>{\centering\arraybackslash}p{1em}}
\usepackage{booktabs}
\usepackage{makecell}
\usepackage{caption}
\usepackage{longtable}
\usepackage{threeparttable, tablefootnote}
\usepackage{lscape}
\usepackage[dvipsnames]{xcolor}
\usepackage{graphicx}

\journal{Machine Learning with Applications}

\newcommand{\chatgpt}{ChatGPT}
\newcommand{\genai}{Generative AI}

\newcommand{\review}[1]{#1}

\begin{document}

\begin{frontmatter}



\title{Benefits and Risks of Using ChatGPT as a Support Tool for Teaching in Computer Science}


\author[inst1]{Yaiza Aragonés-Soria\corref{contrib}}
\ead{yaiza.aragonessoria@constructor.com}

\author[inst1, emailJ]{Julia Kotovich \corref{contrib}}
\ead{Julia.Kotovich@constructor.org}

\author[inst1,inst2, emailC]{Chitsutha Soomlek \corref{contrib}}
\ead{chitsutha@kku.ac.th}

\author[inst1, emailM]{Manuel Oriol\corref{corauthor}}
\ead{mo@constructor.org}

\affiliation[inst1]{organization={Constructor Institute},
            addressline={Rheinweg 9}, 
            city={Schaffhausen},
            postcode={8200}, 
            country={Switzerland}}

\affiliation[inst2]{organization={Department of Computer Science},
            addressline={College of Computing, Khon Kaen University}, 
            city={Khon Kaen},
            postcode={40002}, 
            country={Thailand}}

\cortext[corauthor]{Corresponding author}
\cortext[contrib]{Authors contributed equally}

\begin{abstract}
Upon release, ChatGPT3.5 shocked the software engineering community by its ability to generate answers to specialized questions about coding.
Immediately, many educators wondered if it was possible to use the chatbot as a support tool that helps students answer their programming questions.

This article evaluates this possibility at three levels: fundamental Computer Science knowledge (basic algorithms and data structures), core competency (design patterns), and advanced knowledge (quantum computing). 
In each case, we ask normalized questions several times to ChatGPT3.5, then look at the correctness of answers, and finally check if this creates issues.

The main result is that the performances of ChatGPT3.5 degrades drastically as the specialization of the domain increases: for basic algorithms it returns answers that are almost always correct, for design patterns the generated code contains many code smells and is generally of low quality, but it is still sometimes able to fix it (if asked), and for quantum computing it is often blatantly wrong.
\end{abstract}

\begin{graphicalabstract}
\includegraphics{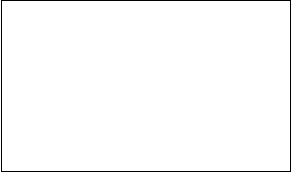}
\end{graphicalabstract}

\begin{highlights}
\item \chatgpt{} can potentially be used for fundamental-level coding in computer science.
\item The generated code and test code are not ready-to-use and exposed to quality issues. 
\item \chatgpt{} is not reliable when asked about quantum computing.
\end{highlights}


\end{frontmatter}


\section{Introduction}\label{sec:intro}
Answering questions, writing emails and essays, creating poems, composing music or solving math problems are only some tasks that chatbots such as Chat Generative Pretrained Transformer (\chatgpt{}) can perform. 
This new technology promises to change our society, especially in education and research~\citep{murugesan2023, damian2023, yilmaz2023, lim2023}. 
Both fields need to adapt to the existence of chatbots, such that they do not become a threatening tool but a support for sharing knowledge. 
The first step for adaptation is to recognise that the change is happening and gather knowledge about its strengths and limitations to take advantage of it. 
It is important that the education and scientific communities become aware of the capabilities of \genai{}.

Since the appearance of \chatgpt{}, the most famous chatbot to date, several studies have tested its capabilities. \citet{PhysicsQuestionsChatGPT}
shows that \chatgpt{} exhibits satisfactory performance when asked formulaic physics problems, which required assuming relevant parameters and writing correct code. 
Its calculation ability and capacity to answer conceptual questions are however limited.

This article explores the benefits and risks of using \chatgpt{} as a support tool for teaching in the context of computer science (CS). 
The study considers three categories: fundamental knowledge (taught to first-year students in computer science), core competency (taught to students finishing a three-year bachelor in computer science), and advanced topics (only taught at the masters level) in computing-related curriculum. 
The approach is to take one course in each category and evaluate how well \chatgpt{} is answering requests. Questions at each level are asked several times to observe the variation in answers. Answers are then categorized as \emph{correct} (complete and accurate), \emph{incomplete} (true but lacking some information), or \emph{incorrect} (containing false information).

The evaluation of fundamental knowledge is a continuation of the work from \citet{fisee23Kotovich}, which showed that \chatgpt{} is answering correctly simple coding assignments. 
The rest of the article is brand new.

For core competency, the reliability, benefits, and risks of using \chatgpt{} to support students on design patterns, anti-patterns, code quality, and software testing is less clear: the present article shows that the resulting code exhibits poor quality, even if most answers are technically correct.

For advanced topics, \chatgpt{} is simply not good enough: it does not answer even simple quantum computing problems based on one-qubit and two-qubit quantum gates.

Whereas our study investigates the use of \chatgpt{}3.5 for teaching computer science, its insights are applicable beyond the educational realm. Lessons learned, especially regarding challenges and benefits, are also relevant for software development. 

Section \ref{sec:methodology} describes the methodology used to test the capabilities of \chatgpt{}. 
Sections \ref{sec:fundmental}, \ref{sec:patterns} and \ref{sec:quantum}, apply the methodology to respectively test \chatgpt{} on fundamental knowledge, core competency, and an advanced topic in computer science. 
Each section first explains the concrete details of the methodology for each topic,and then presents and analyses the results of the experiments. 
Section \ref{sec:implications} present a few conclusions and exercises for students to warn them and make them understand the limitations of ChatGPT and similar systems. 
Section \ref{sec:related_work} summarizes related work. 
Section \ref{sec:conc} presents conclusions and outlook.

\section{Research Methodology}\label{sec:methodology}
This section outlines the general methodology to evaluate the performances of \chatgpt{}. 
Each subsequent section provides further details of the specific methodology at each level: fundamental knowledge, core competency, and advanced topics.

We first select standard questions for each level, we then ask \chatgpt{} version 3.5 each question three times. 
Although \chatgpt{} tends to repeat some sentences, it never gives the exact same answer. 
In case an answer, or part of it, is unclear, we ask \chatgpt{} for a clarification in the same conversation.


All answers are evaluated and classified individually according to the following categories:
\begin{itemize}
    \item \emph{correct answer} - when all information given by \chatgpt{} is true and complete,
    \item \emph{incomplete answer} - when all information given by \chatgpt{} is true but incomplete,
    \item \emph{incorrect answer} - when any information given by \chatgpt{} is false.
\end{itemize}

Finally, to make a quantitative analysis of the answers, the articles reports the percentages of \emph{correct} answers, an \emph{incomplete} answers and an \emph{incorrect} answers. 

All dialogues with \chatgpt{} are stored in a Github repository~\citep{JuliasRepo} to support repeatable and reproducible research.

Several considerations influenced the choice of selecting ChatGPT 3.5 as the focal point of our study. 
One primary factor is the accessibility and affordability for students. 
ChatGPT4, being a more advanced model, comes with a higher cost, potentially limiting its availability to students with constrained budgets. \review{One of the authors reached out to OpenAI to request an institutional subscription but was declined due to high demand at the time. During this period, the company was only permitting individual subscriptions. Additionally, the infrastructure necessary to accommodate large-scale demand for ChatGPT4 appears to be limited, impeding its extensive use in educational environments. This limitation raises concerns regarding fairness and inclusion within classrooms.}  
ChatGPT 3.5, being the most famous iteration in the ChatGPT series, has gained widespread recognition and usage, making it a more practical choice for integration into educational environments. 
Its reputation as the most human-like chatbot makes it a compelling candidate for supporting natural language interactions in teaching and research contexts. 
In our opinion, the combination of its prominence, human-like conversational abilities, and relative affordability makes ChatGPT 3.5 a suitable and pragmatic choice for exploring the potential benefits and risks associated with utilizing generative AI in computer science education and research.

\section{Fundamental Knowledge in Computer Science}\label{sec:fundmental}
Sorting algorithms and data structures are fundamental concepts that all developers should understand and be able to implement. 
With the appearance of \chatgpt{}, one may argue that memorizing the specific code to implement these algorithms or data structures is no longer essential. 
Instead, developers could rely on \chatgpt{} as a supporting tool to generate the desired code snippet. 
This section, evaluates if this scenario is realistic and determines whether students and developers can rely on \chatgpt{}'s abilities to code sorting algorithms and data structures.

\subsection{Experiment}\label{sec:high}
The experimenter asks \chatgpt{} to generate a Python piece of code for a concrete sorting algorithm or data structure. 
We chose Python as a target language because it is taught at many school and universitiess. 
Python is also the most popular general-purpose programming language on~\citet{StackOverflow}.

The algorithms selected for the test correspond to the most used data structure and sorting algorithms presented on~\citet{bigocheatsheet}, which appears as the first result on a Google search that describes ``algorithms complexity'' with a complete list of algorithms (January 2023). 
In that list, there are 13 sorting algorithms and 13 data structures (see Table~\ref{t:fundamental}).


For each data structure and algorithm (except the Array type, which is a base type in Python), ChatGPT is asked to create an implementation in Python using the question \emph{ ``write a python code for X''} (for example \emph{ ``write a python code for Bubble Sort''}).
For each piece of code, ChatGPT provides a code snippet and a short comment such as \emph{``This code sorts an input array 'arr' using the bubble sort algorithm. The algorithm compares each pair of adjacent elements and swaps them in they are if the wrong order. This process is repeated until the array is sorted in ascending order.''}

The resulting (generated) code is copied to an integrated development environment (IDE) to check if it has any errors. 
In case the code can be built and run without an error, the answer is considered \emph{correct}. 
If there are small and easy-to-fix errors, the answer is \emph{incomplete}. 
The answer is \emph{incorrect} when it contains major errors or the code does not align with the algorithm used in the question. 
The percentages in Table~\ref{t:fundamental} are calculated by repeating each question three times.

\begin{figure*}[tb!]
  \centering
  \includegraphics[width=0.7\textwidth]{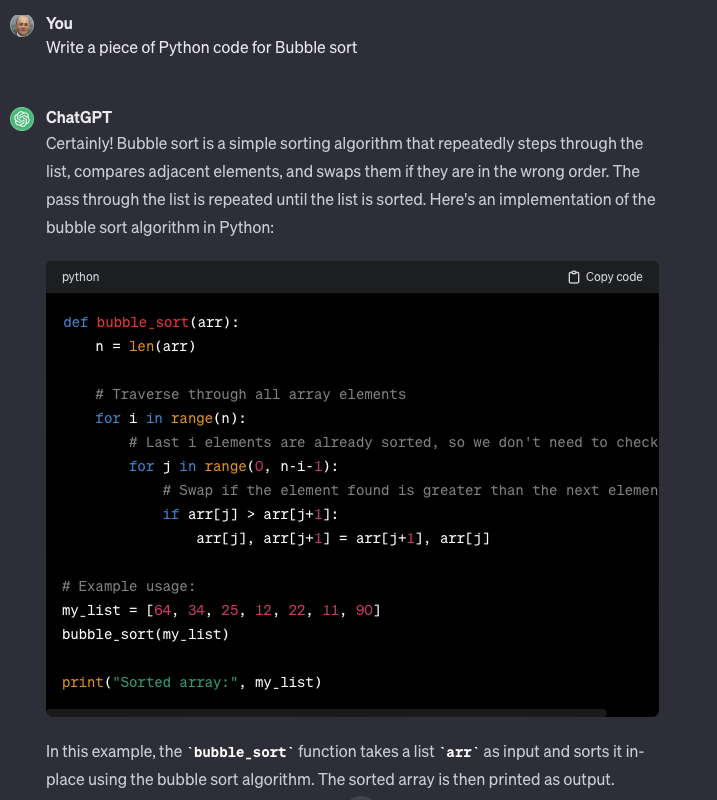}
\caption{First answer of \chatgpt{} on the prompt to code the Bubble Sort Algorithm. }
\label{f:bubble_sort}       
\end{figure*}


\subsection{Results}\label{sec:xp}
\begin{table}[htb!]
    \footnotesize
    \centering    
    \caption{Percentages of correct, incomplete and incorrect Python code generated by \chatgpt{} about sorting algorithms and data structures. Each request is run three times}\label{t:fundamental}
    \begin{tabular}{c*{4}{c}}
    \toprule
      \multirow{2}{*}{Data Structures and Algorithms} & \multicolumn{3}{c}{Percentages} \\
    \cmidrule(lr){2-4}
     & \emph{Correct} & \emph{Incomplete} & \emph{Incorrect} \\
    \midrule
    Quicksort & 0.33 & 0.66 & 0\\
    Mergesort & 1 & 0 & 0 \\ 
    Timsort   & 1 & 0 & 0 \\
    Heapsort  & 1 & 0 & 0 \\
    Bubble Sort & 1 & 0 & 0 \\
    Insertion Sort & 1 & 0 & 0 \\
    Selection Sort & 1 & 0 & 0 \\
    Tree Sort & 1 & 0 & 0 \\
    Shell Sort & 1 & 0 & 0 \\
    Bucket Sort & 1 & 0 & 0 \\
    Radix Sort & 1 & 0 & 0 \\
    Counting Sort & 1 & 0 & 0 \\
    Cubesort & 1 & 0 & 0 \\
    Stack & 0.66 & 0.33 & 0 \\
    Queue & 0.66 & 0.33 & 0 \\
    Singly-Linked List & 0.66 & 0.33 & 0 \\
    Doubly-Linked List & 0 & 1 & 0 \\
    Skip List & 1 & 0 & 0 \\
    Hash Table & 1 & 0 & 0 \\
    Binary Search Tree & 1 & 0 & 0 \\
    Cartesian Tree & 0.66 & 0.33 & 0 \\
    B-Tree & 0.66 & 0.33 & 0 \\ 
    Red-Black Tree & 0.66 & 0.33 & 0 \\ 
    Splay Tree & 0.66 & 0.33 & 0 \\ 
    AVL Tree & 0.66 & 0.33 & 0 \\ 
    KD Tree & 0.66 & 0.33 & 0 \\
    \midrule
    Overall percentages & 82\% & 18\% & 0\%\\
    \bottomrule
    \end{tabular}
\end{table}

\begin{figure*}[tb!]
  \centering
  \includegraphics[width=0.7\textwidth]{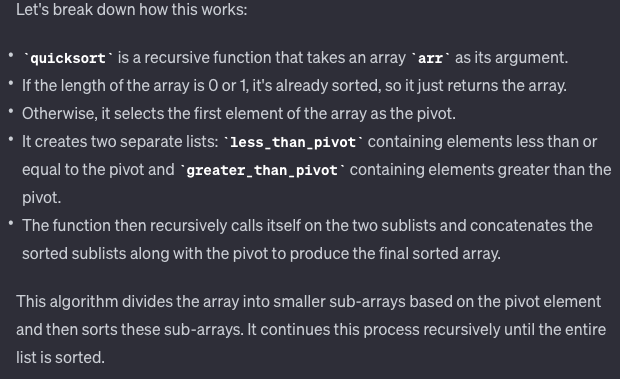}
\caption{Additional explanations answering the question ``You are a computer science instructor with 20 years of experience, and an expert in Python programming. Explain to me how to create a quicksort algorithm in Python.''}
\label{f:explanations}       
\end{figure*}

In 82\% of the cases, the code generated by ChatGPT is \textit{correct}. 
In 18\% of the cases it is incomplete.
It is never \textit{incorrect} (see Table~\ref{t:fundamental}).
The results  confirm that \chatgpt{} can assist students in producing the correct code for the selected data structures and algorithms.
Further to these results, \chatgpt{} can be used as a tutor that explains algorithms (see Figure~\ref{f:explanations}).

For instructors, data structure and algorithms are the pre-requisite and important fundamental knowledge for both basic and more sophisticated programming problems.
If students do not code the algorithm themselves, as with any high-level explanations, answers from \chatgpt{} might create a false sense of understanding that is not validated by a practical implementation.

\chatgpt{} plays an important role in giving examples and guidelines to students, but it also much easier to cheat. 
One might hope that a student may choose to compare their choice of implementation to what \chatgpt{} would offer. 
In this case, \chatgpt{} could assist and accelerate the learning process of the student. 
Maybe more realistically, a student may chose to copy and paste the answer given by \chatgpt{} and only learn superficially.
\subsection{Limitations and Threats to Validity}\label{sec:threat}
The question considered in this section is whether \chatgpt{} is reliable as a support tool for computer science education and research to acquire and validate fundamental knowledge of Computer Science. 
We conclude that \chatgpt{} is indeed a useful support tool at this level giving a correct answer in a high probability.

The main threat to validity to this part of our study is that the algorithms used to test ChatGPT are not representative of all fundamental concepts in CS. 
Indeed, the study only considers a limited number of sorting algorithms and data structures. 
Other topics such as dynamic programming or computational complexity may confirm or refute the conclusions.

\section{Core Competency in Computer Science}\label{sec:patterns}
To develop a high quality software system that meets industrial needs, one must master software design concepts, design patterns, anti-patterns, and software quality. 
The Association for Computing Machinery (ACM) even suggests software design problems, implementation, design software tests, debugging, refactoring an industry computer program, and analysis of trade-offs (e.g., maintainability, efficiency, intellectual property constraints, usability, correctness, graceful failure, efficiency, etc.) as core competencies in curricula recommendations for global computing education and other computing-related disciplines \citep{clear2019computing, gal2020computing}.
This section explores how \chatgpt{} can help with such topics by looking at various design patterns and evaluating the quality of the produced solutions.

A design pattern is a repeatable design solution to a problem and is comprised of a pattern name, a problem, solutions, and consequences~\citep{gamma1995design, freeman2004head}. 
A design pattern is a reusable design template for a common problem found when designing a software. 
Anti-patterns and code smells are the opposite~\citep{fowler1999refactoring,fowler2018refactoring,martin2008clean}. 
They are the solutions that one should avoid. 
Students need to practice thoroughly to be able to analyze and choose a suitable design pattern for a programming problem, and to identify potential design issues. 
A tool such as ChatGPT could be very beneficial to help students reach proficiency.

The world is fascinated by the ability of \chatgpt{} to write pieces of code in the hopes that it accelerates the software development process.
They also hope that it will help students and developers code with good design.
The question is whether it actually produces results of good enough quality.
This research evaluates the benefits and risks of using \chatgpt{} as a support tool for teaching core competency in software engineering. 
This section evaluates the reliability of using \chatgpt{} to assist with design patterns, anti-patterns, code quality, and software testing in the context of computing education. 

The study considers six design patterns: Adapter, Decorator, Observer, Mediator, Singleton, and Abstract Factory.
These patterns are representatives from the three main categories: the structural patterns, creational patterns, and behavior patterns~\citep{gamma1995design}.
The study evaluates Java code, refactored code, JUnit test code, and corresponding answers generated by \chatgpt{}3.5. 

These patterns are also the most commonly used design patterns in open source Java projects~\citep{freeman2004head, hahsler2003quantitative, tsantalis2006design}. 
This limits the exploration space to smaller programs rather than to large/enterprise software systems.
This matches the size of undergraduate-level programming's problems.
The generated source code is then tested and runs on IntelliJ IDEA 2023.1.2 (Ultimate Edition)~\citep{intellij} and the quality of the generated source code is evaluated by well-recognized tools: PMD IntelliJ IDEA plugin~\citep{pmdwebsite}, MetricsTree~\citep{metricstree}, and MetricsReloaded~\citep{metricsreloaded}, as well as by a human expert. 

\subsection{Experiment}\label{sec:patterns_exp}
\begin{figure*}[tb!]
  \centering
  \includegraphics[width=0.7\textwidth]{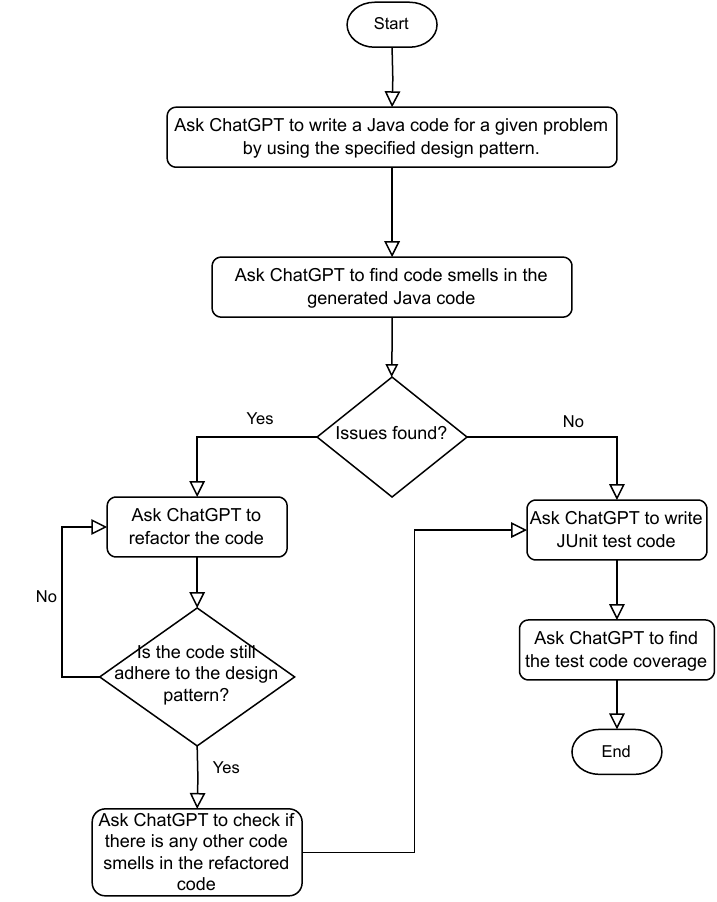}
\caption{Questioning process and data collection}
\label{fig1:flow}       
\end{figure*}

The experiments consist in the workflow presented in Figure~\ref{fig1:flow}.  
The intent is to mimic the approach that a student would take when using \chatgpt{}.
The experiment then evaluates whether \chatgpt{} is a reliable support tool when it is asked to generate a piece of code for a particular design pattern, find design and implementation flaws (i.e., code smells, violations of design concepts, best practice violations, and maintainability issues) in the generated code, refactor the code, and write a unit test for the code. 

Figure~\ref{fig1:flow} illustrates the flow of the questioning process. 
The experimenter starts by asking \chatgpt{} to implement a selected design pattern in the Java programming language for a computer programming problem. 
Then, she asks \chatgpt{} to find code smells in the code. 
If a flaw is found, she then asks \chatgpt{} to refactor the code. 
After that, \chatgpt{} is asked if there are any other code smells in the refactored code. 

Sometimes, the refactored code no longer matches the given design pattern. 
In this case, the experimenter asks \chatgpt{} to check if the code is still conform to the pattern. 
At this step, in some cases, \chatgpt{} may give a new version of the refactored code and/or a descriptive answer. 
In case the code is not given, partially given, or it is still possible to refactor the code further while adhering to the concept of the design pattern, the experimenter asks the \chatgpt{} to refactor again to obtain a more completed version of the refactored code. 

For the final code, the experimenter asks \chatgpt{} to write a unit test in JUnit. 
Finally, the experimenter asks \chatgpt{} to find the test coverage of the generated JUnit tests. 
The experimenter then inspects the design and the structure of code, create a Java project from all of the generated codes, and verifies if the code can run. 
She finally measures the quality of generated code, refactored code, and test code by using software metrics and code analysis tools. 
PMD IntelliJ IDEA plugin~\citep{pmdwebsite} and MetricsTree~\citep{metricstree} identify code smells, Java's best practice violations, and maintainability issues, and measure a wide range of software metrics. Other sets of software metrics are also collected by using MetricsReloaded~\citep{metricsreloaded}.

The information then allows the human experts to verify the answers together with the design diagram generated from the source code and to confirm the quality of the generated code. 
Note that a series of questions formulated for each design pattern are repeated three times to ensure the validity of our experiments.   
\subsection{Results}\label{sec:patterns_results}
At first, the experimenter checks if the generated code matches to the specified pattern, by comparing the design diagram of the code and the structure of the code to the design pattern, and by examining the implementation. 
Then, she checks if the code can run by creating a Java project in IntelliJ IDEA with initial setup. 
If the code cannot run, she examines the errors. 
When the errors require minimal configuration and changes, e.g., setting class path and importing libraries and packages, are needed, the experimenter configures and changes the code accordingly to explore if the code can run. 
In case there are further errors, i.e., build errors and run-time errors, then there is no further change to the code. 
For the generated code, refactored code, and test code, we calculated the percentages of Java code that is ready-to-use, requires minimal changes before it can run, or can not run at all after the configuration and changes have been made. 
Table~\ref{t:pattern_prob_run} shows the results. 

\begin{table}[htb!]
    \footnotesize
    \centering    
    \caption{\label{t:pattern_prob_run} Percentages of pieces of Java code generated by \chatgpt{} that is ready-to-use, requires minimal configurations/changes before it can run, or cannot run.}
    \begin{tabular}{l*{5}{c}}
    \toprule
    Design Pattern & Observation & \multicolumn{3}{c}{Percentages} \\
    \cmidrule(lr){3-5}
    & & Ready-to-use & Configuration & Cannot run \\
    & & & Required & \\
    \midrule
                    & Code & $0\%$ & $\bf100\%$ & $0\%$ \\
    Adapter         & Refactored code & $0\%$ & $\bf83\%$ & $17\%$\\
                    & Test code & $0\%$ & $33\%$ & $\bf67\%$ \\
    \midrule
                    & Code & $0\%$ & $\bf100\%$ & $0\%$ \\
    Decorator       & Refactored code & $0\%$ & $\bf100\%$ & $0\%$ \\
                    & Test code & $0\%$ & $\bf67\%$ & $33\%$ \\
    \midrule
                    & Code & $0\%$ & $\bf100\%$ & $0\%$ \\
    Observer        & Refactored code & $0\%$ & $33\%$ & $\bf67\%$  \\
                    & Test code & $0\%$ & $0\%$ & $\bf100\%$ \\
    \midrule
                    & Code & $0\%$ & $33\%$ & $\bf67\%$ \\
    Mediator        & Refactored code & $0\%$ & $33\%$ & $\bf67\%$\\
                    & Test code & $0\%$ & $0\%$ & $\bf100\%$ \\
    \midrule
                    & Code & $0\%$ & $\bf100\%$ & $0\%$  \\
    Singleton       & Refactored code & $0\%$ & $\bf100\%$ & $0\%$  \\
                    & Test code & $0\%$ & $0\%$ & $\bf100\%$  \\
    \midrule
                    & Code & $0\%$ & $\bf100\%$ & $0\%$   \\
    Abstract Factory& Refactored code & $0\%$ & $40\%$ & $\bf60\%$ \\
                    & Test code & $0\%$ & $33\%$ & $\bf67\%$  \\
    \midrule
                    & Code & $0\%$ & $\bf89\%$ & $11\%$ \\
    \textbf{Overall percentages}  & Refactored code & $0\%$ & $\bf65\%$ & $35\%$   \\
                    & Test code & $0\%$ & $28\%$ & $\bf82\%$  \\
    \bottomrule
    \end{tabular}
\end{table}

\begin{figure*}[tb!]
  \centering
  \includegraphics[width=1.0\textwidth]{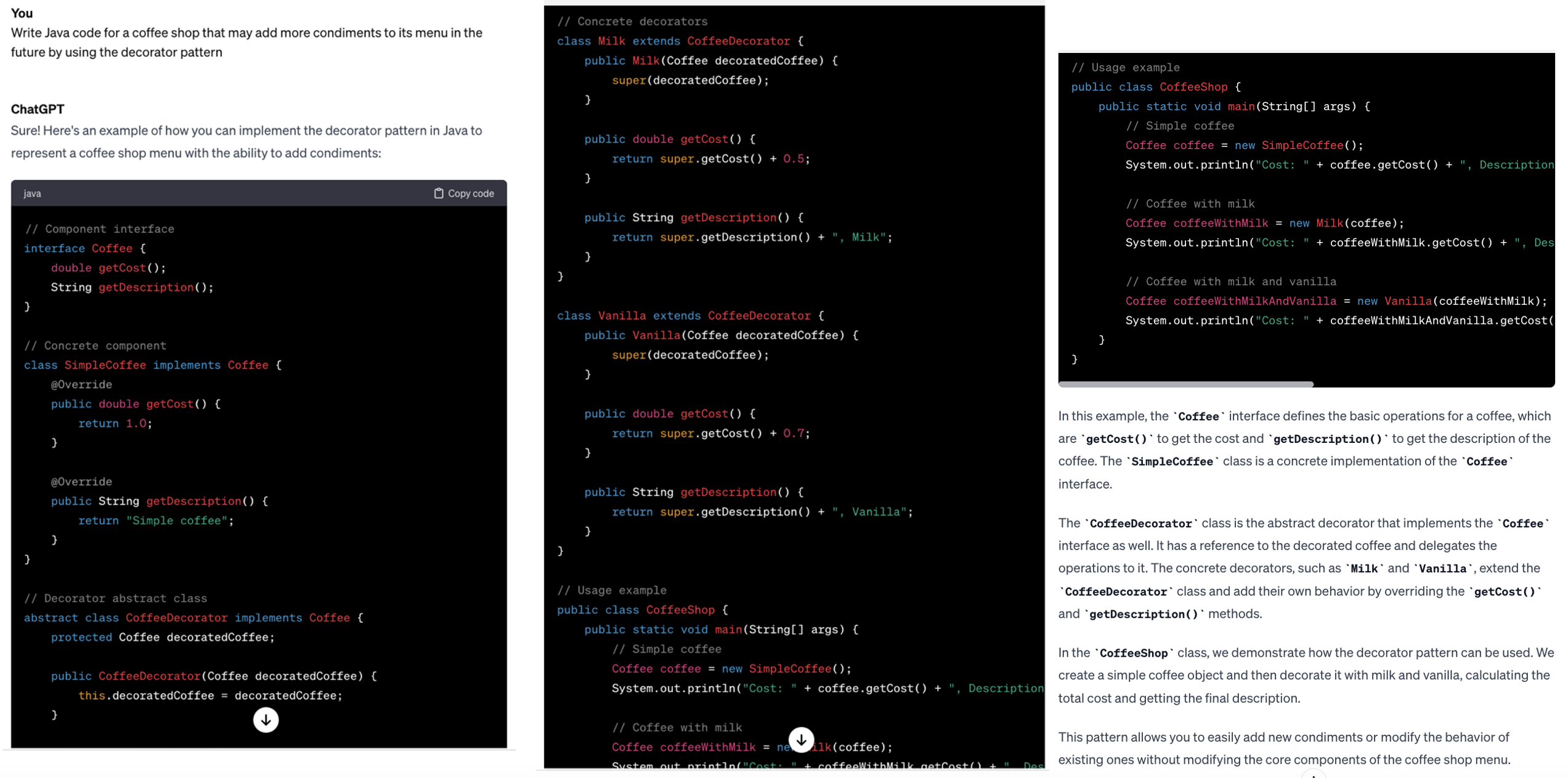}
\caption{Example of code snippet generated by ChatGPT. }
\label{f:snippet_decorator}       
\end{figure*}

The experimental results confirm that the generated code, refactored code, and test code are not ready to use. 
The major reason is that \chatgpt{} always writes everything in one single file. 
When writing a Java program, it is important to separate Java classes into files with names matching the class name, especially when the class is public. 
Note that it is possible to have multiple classes in a Java file. 
A package in Java is commonly used to organize the related classes. 
These are the parts of Java specific best practices since each class can be executed separately to confirm their functionalities and to support maintainability. Figure ~\ref{f:snippet_decorator} gives an example of a code snippet of the Decorator Pattern generated by ChatGPT.

When creating a Java project that contains the generated code, it is therefore needed to create also a package and a \emph{test} package in the project, separate classes into files, change the code for importing classes from related packages, and configure as required to ensure that the code runs. 
During the experiment, the experimenter does not change the design of the program and also does not modify any logic inside the code. 
As reported in Table~\ref{t:pattern_prob_run}, the majority of generated answers that cannot run are regarding test code. 
There are two major causes for this: (1) assertion on non-existing methods and (2) creation of an object from a non-existing class. 
Both cases lead to build errors and therefore the tests cannot be executed.

\begin{table}[htb!]
    \small
    \centering    
    \caption{Percentages of to obtain a \emph{correct} answer, an \emph{incomplete} answer, or an \emph{incorrect} answer when asking \chatgpt{} to generate a Java piece of code and its test code.}\label{t:pattern_prob_correct}
    \begin{tabular}{l*{5}{c}}
    \toprule
    Design Pattern & Observation & \multicolumn{3}{c}{Percentages} \\
    \cmidrule(lr){3-5}
    & & \emph{Correct} & \emph{Incomplete} & \emph{Incorrect} \\
    \midrule
                    & Code  & $\bf100\%$ & $0\%$ & $0\%$  \\
    Adapter         & Refactored code  & $\bf33\%$ & $\bf33\%$ & $\bf33\%$\\ 
                    & Test code  & $0\%$ & $\bf100\%$ & $0\%$ \\
                    & Test coverage & $0\%$ & $0\%$ & $\bf100\%$ \\
    \midrule
                    & Code  & $\bf100\%$ & $0\%$ & $0\%$  \\
    Decorator       & Refactored code  & $\bf67\%$ & $33\%$ & $0\%$ \\
                    & Test code  & $33\%$ & $\bf67\%$ & $0\%$ \\
                    & Test coverage & $0\%$ & $0\%$ & $\bf100\%$ \\
    \midrule
                    & Code  & $\bf100\%$ & $0\%$ & $0\%$  \\
    Observer        & Refactored code  & $0\%$ & $\bf100\%$ & $0\%$ \\
                    & Test code  & $0\%$ & $\bf100\%$ & $0\%$ \\
                    & Test coverage & $0\%$ & $0\%$ & $\bf100\%$ \\
    \midrule
                    & Code  & $33\%$ & $\bf67\%$ & $0\%$ \\
    Mediator        & Refactored code  & $0\%$ & $33\%$ & $\bf67\%$ \\
                    & Test code  & $0\%$ & $\bf100\%$ & $0\%$ \\
                    & Test coverage & $0\%$ & $0\%$ & $\bf100\%$ \\
    \midrule
                    & Code  & $\bf100\%$ & $0\%$ & $0\%$  \\
    Singleton       & Refactored code  & $\bf75\%$ & $0\%$ & $25\%$ \\
                    & Test code  & $0\%$ & $\bf1.00$ & $0\%$ \\
                    & Test coverage & $0\%$ & $0\%$ & $\bf100\%$ \\
    \midrule
                    & Code  & $\bf100\%$ & $0\%$ & $0\%$  \\
    Abstract Factory& Refactored code  & $40\%$ & $0\%$ & $\bf60\%$ \\
                    & Test code  & $0\%$ & $\bf67\%$ & $33\%$ \\
                    & Test coverage & $0\%$ & $0\%$ & $\bf100\%$ \\
    \midrule
                    &Code & $\bf89\%$ & $11\%$ & $0\%$   \\
    \textbf{Overall percentages}  & Refactored code & $38\%$ & $29\%$ & $33\%$    \\
                    & Test code & $6\%$ & $\bf88\%$ & $6\%$   \\
                    & Test coverage & $0\%$ & $0\%$ & $\bf100\%$  \\
    \bottomrule
    \end{tabular}
\end{table}

To analyse further the answers, the experimenter also computes the percentages of \emph{correct} answers, \emph{incomplete} answers, and \emph{incorrect} answers. The criteria used to evaluate a \chatgpt{}'s response are as follows:
\begin{itemize}
    \item {\verb|Code|}: A generated piece of code is \emph{correct} when it conforms to the requested design pattern and the functionalities of the implementation matches to the given programming problem. When a piece of code aligns either to the pattern or the problem, the code is \emph{incomplete}. A piece of code is \emph{incorrect} when it neither matches the specified design pattern nor the problem.
    \item {\verb|Refactored code|}: A refactored piece of code is \emph{correct} when it conforms to the given design pattern, the functionalities of the implementation is matched to the given programming problem, and at least one code smell/design flaw previously found by \chatgpt{} itself is removed. When a refactored piece of code aligns to the design pattern, it matches the programming problem, or at least one code smell/design flaw is improved, the refactored code is \emph{incomplete}. A refactored piece of code is \emph{incorrect} when it does not conform to the specified design pattern, does not provide a functionality to solve the given problem, and does not improve any of the code smells/design flaws found in the previous step.
    \item {\verb|Test code|}: JUnit test code is \emph{correct} when the test cases match the functionalities of the final version of the refactored code and there is at least one test case for each method in a class. Test code is \emph{incomplete} when a method or a functionality of the refactored code is not tested. Test code is \emph{incorrect} when a generated test case is not related to the functionalities of the refactored code.
    \item {\verb|Test coverage|}: A \chatgpt{}'s response is \emph{correct} when all of the measurement results match the test coverage calculated by IntelliJ IDEA for the same testing coverage techniques. An answer is \emph{incomplete} when one of the produced test coverage results does not comply to the test coverage calculated by IntelliJ IDEA for the same testing coverage technique. The response is \emph{incorrect} when nothing matches the results obtained from IntelliJ IDEA for the same testing coverage technique. 
\end{itemize}

Table~\ref{t:pattern_prob_correct} shows the percentages of \emph{correct} answers, \emph{incomplete} answers, and \emph{incorrect} answers. 
Very often, \chatgpt{} produces \emph{correct} results for most of the design patterns, except for the mediator pattern. 
When it refactors the code, the improved code does usually no longer strictly adhere to the given design pattern. 
An experienced developer may choose not to strictly follow the chosen design pattern, but rather modify the design and the code to achieve the design goals while considering potential trade-offs.
In that respect \chatgpt{} is acting similarly to a developer.

For students it is important to have the ability to verify and validate the design and code suggested by \chatgpt{}, which can be considered as both benefits and risks of using it as a support tool for learning.
On the bright side, a student gains more knowledge and insight from examining the response thoroughly. 
A student can also compare the response to their answers and other sophisticated resources to learn more. 
Unfortunately, if a student is not vigilant enough to the \emph{incorrect} or \emph{incomplete} answers of \chatgpt{}, the student might trust unreliable knowledge resource. 

\begin{figure*}[tb!]
  \centering
  \includegraphics[width=0.95\textwidth]{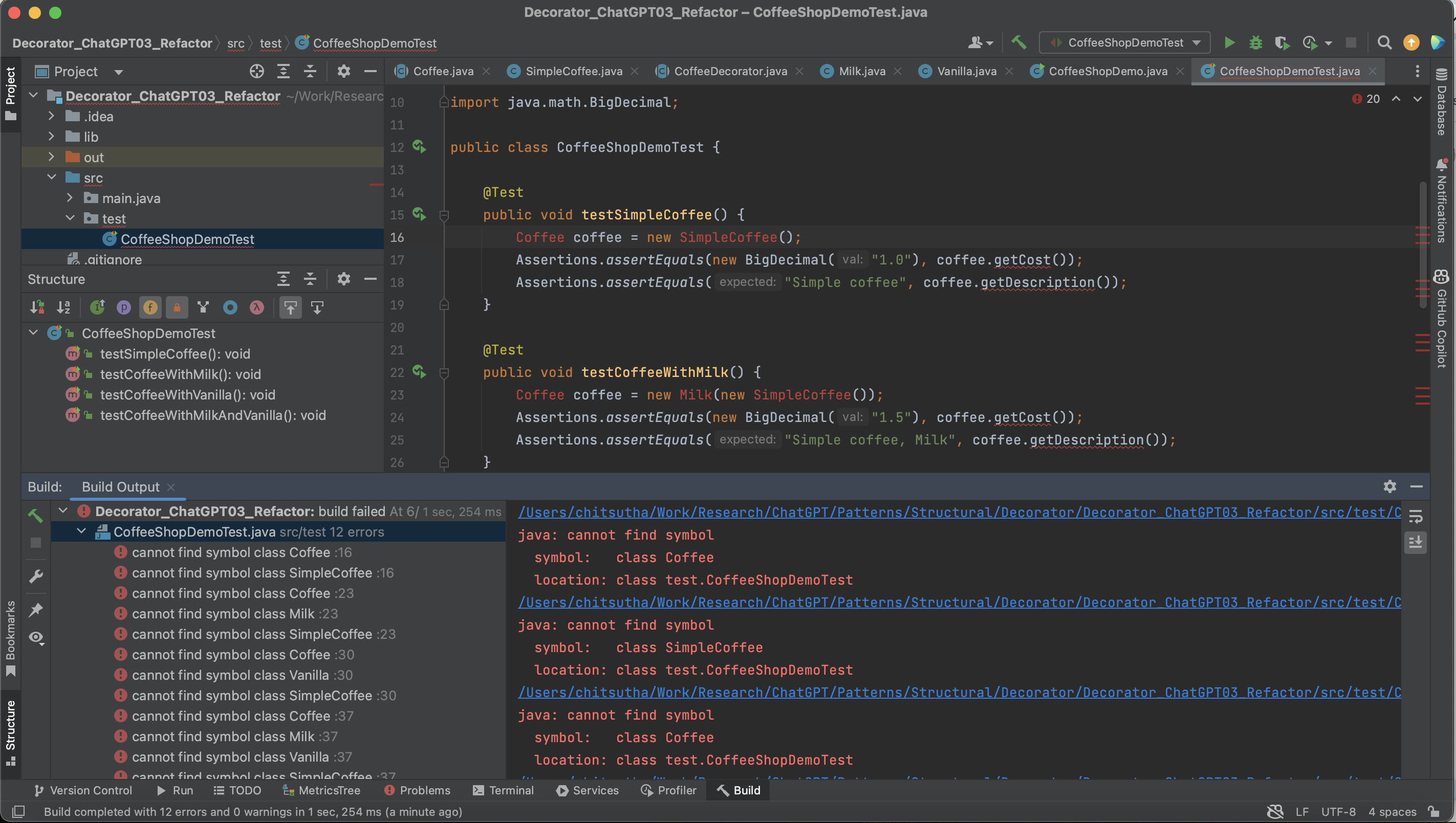}
\caption{Partially correct test code. The test code can be run after those errors are resolved.}
\label{f:testerror}       
\end{figure*}

\begin{figure*}[tb!]
  \centering
  \includegraphics[width=0.57\textwidth]{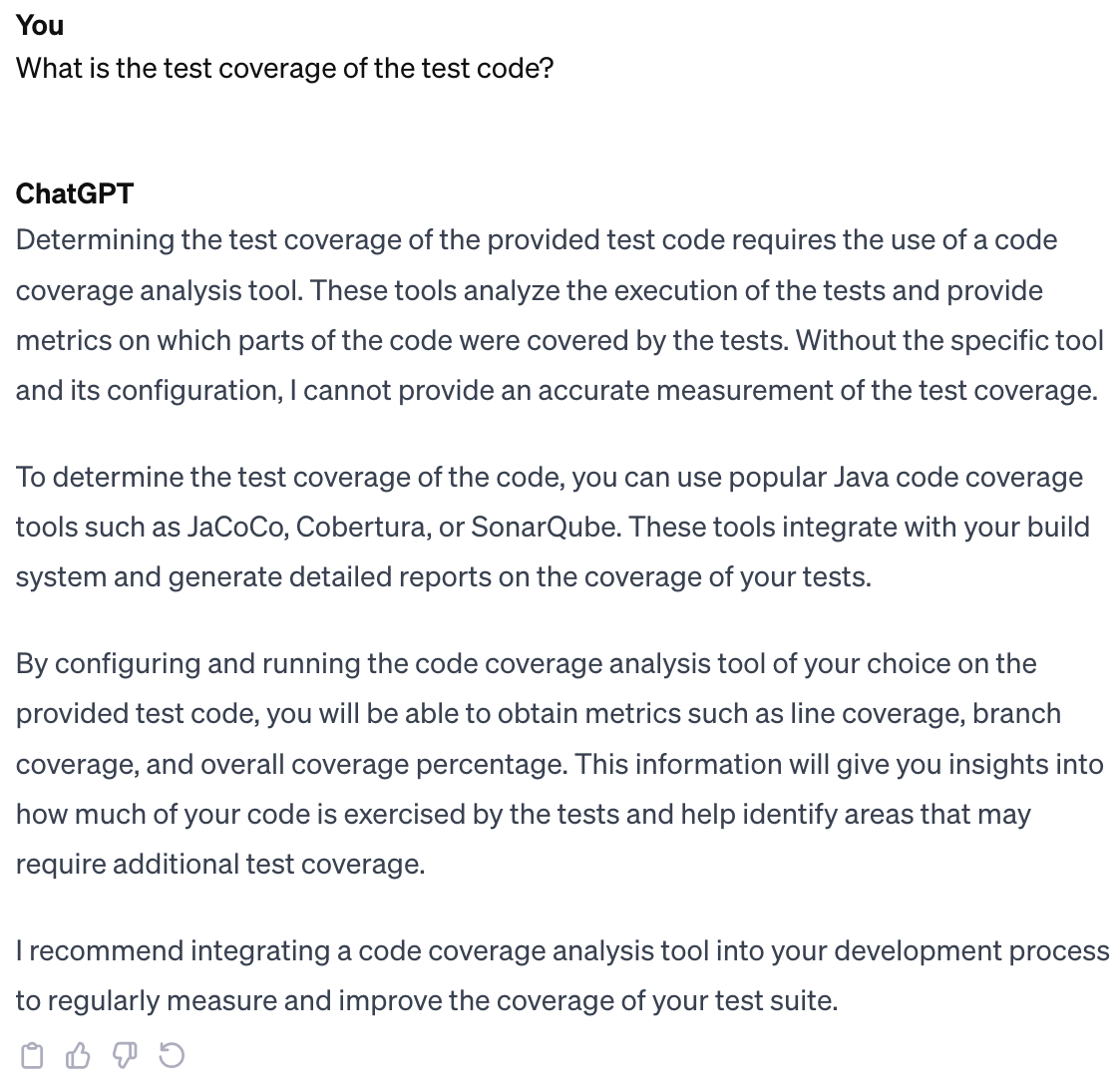}
\caption{Response from ChatGPT3.5 when asking to find test coverage. }
\label{f:testcoverage}       
\end{figure*}

For test code, \chatgpt{} can summarize the main functionalities of some source code, but it does not work well in designing a test case or generating test code. 
As mentioned previously, most of the test code cannot run because of build failures, as shown in Figure~\ref{f:testerror}. The test code in the given example can be run after those errors are resolved. 
The test cases also do not cover all of the functionalities of the relevant code. 
In few cases, a generated test case is not related to a functionality of the code.
In addition, since \chatgpt{} version 3.5 is not connected to a code coverage tool (even if it can suggest the appropriate tool), it cannot compute the correct test coverage results (see Figure~\ref{f:testcoverage} for ChatGPT's reponse).

\begin{table}[htb!]
    \begin{threeparttable}
    \caption{Number of the quality issues found in the generated code.} \label{t:pattern_pmd}
    \tiny
    \centering 
    \begin{tabular}{l*{10}{c}}
    \toprule
    Design & Observation & \multicolumn{7}{c}{Number of Code Quality Issues Found by PMD} & Total\\
    Pattern & \\
    \cmidrule(lr){3-9}
    & &  Best & Coding & Design & Error & Doc. & Perf. & Thread\\
    & & practice & style &  & prone &  &  &\\
    \midrule
            & Code\#1  & $1$ & $11$ & $2$ & $0$ & $12$ & $0$ & $0$ & $26$\\
            & Refactor\#1  & $1$ & $10$ & $3$ & $0$ & $13$ & $0$ & $0$ & $\bf27$\\
            & Test\#1  & $0$ & $4$ & $0$ & $1$ & $0$ & $0$ & $0$ & $5$\\
            & Code\#2  & $2$ & $12$ & $3$ & $0$ & $12$ & $0$ & $0$ & $29$\\
    Adapter & Refactor\#2  & $4$ & $9$ & $3$ & $0$ & $16$ & $0$ & $0$ & $\bf32$\\
            & Test\#2  & $3$ & $8$ & $0$ & $1$ & $0$ & $0$ & $0$ & $12$\\
            & Code\#3  & $1$ & $10$ & $2$ & $0$ & $12$ & $0$ & $0$ & $\bf25$\\
            & Refactor\#3  & $1$ & $10$ & $2$ & $0$ & $12$ & $0$ & $0$ & $\bf25$\\
            & Test\#3  & $1$ & $8$ & $0$ & $0$ & $1$ & $0$ & $0$ & $10$\\
    \midrule
             & Code\#1  & $9$ & $11$ & $1$ & $0$ & $19$ & $0$ & $0$ & $\bf40$\\
             & Refactor\#1  & $1$ & $12$ & $1$ & $0$ & $13$ & $0$ & $0$ & $27$\\
             & Test\#1  & $9$ & $8$ & $0$ & $1$ & $0$ & $0$ & $0$ & $18$\\
             & Code\#2  & $2$ & $7$ & $7$ & $0$ & $19$ & $0$ & $0$ & $\bf35$\\
    Decorator& Refactor\#2  & $2$ & $7$ & $4$ & $0$ & $19$ & $0$ & $0$ & $32$\\
             & Test\#2  & $9$ & $8$ & $0$ & $1$ & $1$ & $0$ & $0$ & $19$\\
             & Code\#3  & $2$ & $8$ & $6$ & $0$ & $19$ & $0$ & $0$ & $\bf35$\\
             & Refactor\#3  & $2$ & $10$ & $1$ & $0$ & $13$ & $0$ & $0$ & $26$\\
             & Test\#3  & $0$ & $5$ & $0$ & $0$ & $5$ & $0$ & $0$ & $10$\\
    \midrule
             & Code\#1  & $5$ & $12$ & $4$ & $0$ & $17$ & $0$ & $0$ & $38$\\
             & Refactor\#1  & $11$ & $12$ & $7$ & $0$ & $22$ & $0$ & $0$ & $\bf52$\\
             & Test\#1  & $12$ & $8$ & $0$ & $0$ & $10$ & $0$ & $0$ & $30$\\
             & Code\#2  & $1$ & $14$ & $3$ & $0$ & $14$ & $0$ & $0$ & $32$\\
    Observer & Refactor\#2  & $1$ & $15$ & $3$ & $1$ & $14$ & $1$ & $0$ & $\bf35$\\
             & Test\#2  & $1$ & $2$ & $0$ & $5$ & $0$ & $0$ & $0$ &$8$\\
             & Code\#3  & $3$ & $13$ & $3$ & $0$ & $15$ & $0$ & $0$ & $34$\\
             & Refactor\#3  & $3$ & $13$ & $4$ & $0$ & $15$ & $0$ & $0$ & $\bf35$\\
             & Test\#3  & $7$ & $5$ & $0$ & $0$ & $1$ & $0$ & $0$ & $13$\\
    \midrule
             & Code\#1  & $3$ & $19$ & $2$ & $1$ & $19$ & $0$ & $0$ & $44$\\
             & Refactor\#1  & $3$ & $21$ & $2$ & $1$ & $19$ & $0$ & $0$ & $\bf46$\\
             & Test\#1  & $3$ & $3$ & $1$ & $0$ & $9$ & $0$ & $0$ & $16$\\
             & Code\#2  & $5$ & $18$ & $2$ & $1$ & $20$ & $0$ & $0$ & $46$\\
    Mediator & Refactor\#2  & $5$ & $17$ & $4$ & $1$ & $20$ & $0$ & $0$ & $\bf47$\\
             & Test\#2  & $1$ & $2$ & $1$ & $0$ & $4$ & $0$ & $0$ & $8$\\
             & Code\#3  & $4$ & $19$ & $6$ & $1$ & $19$ & $0$ & $0$ & $49$\\
             & Refactor\#3  & $4$ & $22$ & $2$ & $3$ & $21$ & $0$ & $0$ & $\bf52$\\
             & Test\#3  & $1$ & $3$ & $1$ & $2$ & $5$ & $0$ & $0$ & $12$\\
    \midrule
             & Code\#1  & $3$ & $6$ & $5$ & $2$ & $8$ & $1$ & $2$ & $\bf27$\\
             & Refactor\#1  & $0$ & $6$ & $5$ & $2$ & $7$ & $1$ & $0$ & $21$\\
             & Test\#1  & $2$ & $4$ & $4$ & $0$ & $5$ & $1$ & $0$ & $16$\\
             & Code\#2  & $3$ & $6$ & $5$ & $2$ & $8$ & $1$ & $2$ & $27$\\
    Singleton& Refactor\#2  & $0$ & $6$ & $5$ & $2$ & $7$ & $1$ & $0$ & $\bf21$\\
             & Test\#2  & $2$ & $4$ & $4$ & $0$ & $5$ & $1$ & $0$ & $16$\\
             & Code\#3  & $3$ & $6$ & $5$ & $2$ & $8$ & $1$ & $2$ & $27$\\
             & Refactor\#3  & $0$ & $7$ & $9$ & $1$ & $14$ & $1$ & $3$ & $\bf35$\\
             & Test\#3  & $3$ & $4$ & $2$ & $3$ & $6$ & $1$ & $0$ &$19$\\
    \midrule
             & Code\#1  & $6$ & $21$ & $2$ & $0$ & $12$ & $0$ & $0$ & $41$\\
             & Refactor\#1  & $8$ & $32$ & $1$ & $0$ & $26$ & $0$ & $0$ & $\bf67$\\
             & Test\#1  & $10$ & $11$ & $0$ & $7$ & $0$ & $0$ & $0$ & $28$\\       
    Abstract & Code\#2  & $20$ & $15$ & $1$ & $0$ & $29$ & $0$ & $0$ & $65$\\
    Factory  & Refactor\#2  & $21$ & $14$ & $1$ & $0$ & $35$ & $0$ & $0$ & $\bf71$\\
             & Test\#2  & $14$ & $7$ & $1$ & $0$ & $8$ & $1$ & $0$ & $31$\\
             & Code\#3  & $8$ & $14$ & $2$ & $0$ & $22$ & $0$ & $0$ & $46$\\
             & Refactor\#3  & $13$ & $16$ & $2$ & $0$ & $31$ & $0$ & $0$ & $\bf62$\\
             & Test\#3  & $18$ & $12$ & $7$ & $0$ & $0$ & $0$ & $0$ & $37$\\
    \bottomrule
    \end{tabular}
    \end{threeparttable}
\end{table} 

\begin{table}[htb!]
    \captionsetup{font=scriptsize}
    \caption{Maintainability index across the Java projects and the latest refactored Java project; where $0-9 = Highly\ maintainable, 10-19 = Moderate\ maintainable, 20-100 = Difficult\ to\ maintain;$ code smells and other software quality issues found.}\label{t:pattern_maintain}
    \scriptsize
    \centering    
    \begin{tabular}{c c c c l}
    \toprule
    \multirow{2}{*}{\begin{tabular}{c}
         Design \\ 
         Pattern
    \end{tabular}} & \multirow{2}{*}{\begin{tabular}{c}
         Observation \\ 
         No.
    \end{tabular}} & \multicolumn{2}{c}{Maintainability Index} & \multicolumn{1}{c}{\multirow{2}{*}{\begin{tabular}{c}
         Code Smells and Other \\ 
         Quality Issues Found
    \end{tabular}}} \\
    \cmidrule(lr){3-4}
     & & Code & Refactored Code & \\
    \midrule
                    & $1$ & $53.8$ & $54.5$ & Low tight class cohesion\\
    Adapter         & $2$ & $52.6$ & $48.9$ & Low tight class cohesion\\
                    & $3$ & $53.4$ & $53.4$ & Low tight class cohesion \\
    \midrule
                    & $1$ & $45.5$ & $45.9$ & Magic values, Duplicated code,  \\
                    &     &    &  & Low tight class cohesion, \\
                    &     &    &  &  High number of children \\
    Decorator       & $2$ & $44.4$ & $44.5$ & Magic values, \\
                    &     &           &           & Low tight class cohesion, \\
                    &     &           &           & High number of children, \\
                    &     &           &           & Low weight of a class\\
                    & $3$ & $45.3$ & $45.4$ & Low tight class cohesion, \\
                    &     &           &           & High number of children, \\
                    &     &           &           & Low weight of a class\\
    \midrule
                    & $1$ & $47.0$ & $43.9$ & Low tight class cohesion \\
    Observer        & $2$ & $45.6$ & $47.9$ & Low tight class cohesion \\
                    & $3$ & $46.3$ & $39.9$ & Long method, \\                
                    &     &           &           & Low tight class cohesion \\  
    \midrule
                    & $1$ & $42.6$ & $41.1$ & Long method \\
                    &     &           &              &  High number of children \\
                    &     &           &              &  Low tight class cohesion \\
    Mediator        & $2$ & $42.4$ & $41.8$ & Long method, \\
                    &     &           &           &  High number of children, \\
                    &     &           &           &  Low tight class cohesion, \\
                    &     &           &           &  High weight of a class \\
                    & $3$ & $42.8$ & $37.5$ & High number of children, \\
                    &     &           &           &  Low tight class cohesion \\
    \midrule
                    & $1$ & $53.1$ & $58.9$ & Magic values,\\              
                    &     &              &           & Java Thread Safety is not used \\
    Singleton       & $2$ & $53.1$ & $58.8$ & Magic values,\\              
                    &     &              &           & Java Thread Safety is not used \\
                    & $3$ & $49.5$ & $44.4$ & Magic values,\\              
                    &     &              &           & Java Thread Safety is not used \\
                    &     &              &           & Low tight class cohesion\\
    \midrule
                    & $1$ & $53.1$ & $36.8$ & Long method, Magic values,\\
                    &     &           &            & Duplicated code, \\
    Abstract        &     &           &            & Low number of children  \\
    Factory         & $2$ & $44.7$ & $42.6$ & Long method, Magic values,\\
                    &     &           &            & Duplicated code \\
                    & $3$ & $41.2$ & $44.3$ & Long method, Magic values,\\
                    &     &           &            & Duplicated code \\
    \bottomrule
    \end{tabular}
\end{table}

A reliable support tool for teaching should be able to support a student to achieve the learning objectives, learning outcomes, and core competencies needed. 
The ability to develop high quality software to meet industrial needs is also essential for a computer science student.

This research also examines the quality issues found in the generated code, refactored code, and test code. 
Table~\ref{t:pattern_pmd} provides the number of quality issues relative to Java rulesets identified by PMD, where \emph{Best practice} = no. of best practice violations, \emph{Coding style} = no. of Java specific coding style violations, \emph{Design} = no. of design issues, \emph{Error prone} = no. of either broken, extremely confusing, or prone to run-time errors constructs, \emph{Doc.} = no. of code documentation issues, \emph{Perf.} = no. of performance issues, and \emph{Thread} = no. of issues relative to multiple threads of execution. 
For detailed source code analyzed results, please refer to our GitHub repository.
Table~\ref{t:pattern_maintain} presents the maintainability index, code smells, and other software quality issues found by the code analysis tools and human experts. 

Although \chatgpt{} was asked to find code smells in the generated code, the experimenter did not compare the obtained responses with other tools. 
There are still open questions on the subjectivity of metrics and code smells, lacking of common definitions, the high false positive rate results, and the reliability of software metrics and thresholds of the metric-based code analysis tools~\citep{soomlek2021automatic,bafandeh2020bad,sharma2018survey}.
As a result it is not possible to rely on the tools to obtain the ground truth.

Although the concept of code smells has been introduced and catalogued for decades, many well-known code analysis tools still adopt the set of metrics and corresponding threshold values from Lanzy \textit{et al.}\citet{lanza2006object}. 
As such, their analyzed results may not represent the characteristics of modern programming languages but it is worth examining the results produced from both \chatgpt{} and code analysis tools to confirm the results indicating that the generated code is not ready to use and still needs to be improved to match the design goals and the selected quality attributes.

In conclusions, there are many benefits and risks in using \chatgpt{} as a support tool for teaching a core competency in computer science. 
Obviously, \chatgpt{} can work with a large number of students at the same time, and can help students from anywhere and at anytime.
According to these experiments and observations, both instructors and students should take the following issues into consideration when deciding to use \chatgpt{} as a support tool for teaching:

\paragraph{Background knowledge in software design and programming is required}
The generated code and test code are not ready-to-use.
To utilize generated code, a student needs some basic knowledge and experience in hands-on computer programming. 
A student should be able to setup and initialize a project, configure corresponding environment variables and settings, import the libraries and packages needed, and resolve errors and warnings so that the generated source code can be executed. 
Moreover, students also need to be very precise about what they want from \chatgpt{}. 
For example, there are various implementation choices for the singleton design pattern. 
If a student would like to employ Java Thread Safety in the code, the keywords \emph{Java Thread Safety} must be included in the question. 
Otherwise, \chatgpt{} can freely answer with any choice of implementations that adheres to the singleton design pattern. 
This implies that background knowledge in software design and computer programming is needed to achieve the best possible results from \chatgpt{}.

The same idea is also applied to code refactoring. 
A user must be very specific with \chatgpt{} on how to refactor the code. 
Otherwise, the resulting code might not match the intent of the user. 
Therefore, a user should have a certain level of knowledge relative to software design, programming, and code refactoring to instruct \chatgpt{} to work towards the right direction.

It is important that educators realize that \genai{} tools are especially useful to write code when programmers know what they are doing.
Since they are still in training, how to bring students to the point where they know what they are doing? 
Our point of view, that would need further evaluations, is that educators should train students to use \genai{} tools first and then train students to test these tools and their limits.
In our educators experience one of the best way to train someone on a topic is to let that person teach the topic.
An interesting idea is to make \genai{} tools that would need to be trained by students to perform properly. 
This would help students understand the limits of \genai{} as well as the topic taught.

\paragraph{The generated test code might not match the intent}
There are many test design techniques that a student could adopt to cover positive test cases, negative test cases, boundary conditions, and error conditions without doing exhaustive testing. 
For experienced testers, they may add more test cases after following the selected test design techniques to increase the exploration space.
In these experiments, it is clear that \chatgpt{} can generate test code with correct syntax, but the test code does not align to the code under test. 
Assertions are made on a non-existing method and an object is created from a non-existing class. 
Sometimes, the generated test code is not related to the functionality of the program. 
Therefore, it is possible for students to use \chatgpt{} to program a test in a chosen programming language; which saves them a large amount of time to study on the syntax of a programming language, but they should have the ability to design a test case and instruct \chatgpt{} on how to write the test code to obtain the expected answers.

In addition, a student should be able to verify that the generated test code is adhere to their test design and be able to measure the test coverage at different levels of granularity. 
Since \chatgpt{} 3.5  does not include a test analysis tool, it cannot measure test coverage and mutation scores, as illustrated in Figure~\ref{f:testcoverage}. \chatgpt{}, however, can guide a student to the suitable tools to perform the task.  

\paragraph{The given explanation may not match to the generated code}
When \chatgpt{} finished generating code, it always gives an explanation or a summary of the code. 
Especially, when we asking \chatgpt{} to refactor code, \chatgpt{} always explains what was changed and how the code was improved. 
The information is helpful for users to identify the changes, to go through the changes, and when comparing the refactored code to the original code. 
It is however possible that the given explanation is \emph{incomplete}. 
Sometimes, parts of the explanation do not match to the refactored code. 
In addition, there are changes that are not mentioned in the explanation. 

Course instructors can use the responses as examples of \emph{correct}, \emph{incomplete}, or \emph{incorrect} solutions. 
The generated answers are a useful resource and support for example-based learning because examples help novices to learn more with less time and effort while maximize learning and training efficiency. 
Using those examples without a guideline from an instructor is however risky because a student could gather wrong information from the \emph{incomplete} and \emph{incorrect} examples. 
Therefore, we recommend a course instructor to use the generated code and its explanation together with correct information and guidelines from the instructor, but we do not encourage to solely rely on \chatgpt{} as a support tool for teaching in the context of Computer Science education.     

\paragraph{Quality of the generated code}
Everyone can code, but not everyone codes well.
For a computer science student, it is important to have the ability to code and develop a high quality software system to meet industrial standards. 
Therefore, design patterns, anti-patterns, best practices, coding standards, naming convention, software metrics, and software quality are core courses. 
By using \chatgpt{} as a support tool for teaching computer science, teachers need to ensure the quality of the generated examples so that students can learn how to produce a high quality computer program through good examples. 
According to our experimental results, there are various issues found in the coding samples, e.g., difficult to maintain, code smells, anti-patterns, violation of best practices, error prone, high coupling, low cohesion, and so on and so forth. 

Sometimes the refactored code is no longer following the design pattern indicated at the beginning of the conversation. 
The worst-case scenario is if a student imitates the coding style. 
Therefore, when using \chatgpt{} as a support tool for teaching, it is important for a course instructor to remind their students to verify the results against the design problems, design principles, and design concepts while considering the trade-offs. 
Students should {\bf never} adopt the solution without checking the generated answers. 

\paragraph{Beware of partially correct and incorrect results}
\chatgpt{} is very helpful during the code review process. 
\chatgpt{} helps looking back at the code and think about our design, code smells, best practice violations, design issues, maintainability issues, naming convention, coding standard used, and documentations. 
Even though the results are not always correct, \chatgpt{} works similarly to a group of developers working on code reviews. 
Subjectivity is important for reviewing properly code, it is needed to rethink and consider about the requirements, the potential solutions, and the trade-offs before we can decide on how to improve the code and finalize our choices. 
ChatGPT plays a good role as a support tool for teaching CS, but it is mandatory for the user of \chatgpt{} to have the ability to verify and validate the results. 

This evaluation confirms that the generated responses include \emph{correct}, \emph{incomplete}, and \emph{incorrect} answers. 
These generated answers therefore should be used with caution, especially by novices.

A human teaching assistant can make use of \chatgpt{} as a support tool for teaching, but their tasks go beyond that. A human teaching assistant plays an important role for students who need extra support to reach their full potential, can encourage students to engage more in a class, and can contribute to both educational, emotional, and social development of a student, each of which is a skill set that an artificial intelligence does not have at the time of writing. 
We encourage a teaching assistant and an instructor to use a \genai{} to automate some tasks and relieve their workload.

\subsection{Limitations and Threats to Validity}\label{sec:patterns_threats}
\paragraph{\chatgpt{} 3.5 architecture is not the only representative of a \genai{}} 
As mentioned previously experiments could use ChatGPT4 and result in different results. 
In addition, other large language models and \genai{} specifically trained to support coding and code completion, e.g., Replit Ghostwriter~\citep{ghostwriter}, GitHub Copilot~\citep{copilot}, Hugging Face~\citep{huggingface}, tabnine~\citep{tabnine}, etc., that were not evaluated in this research. 
One could adopt our research methodology and series of questions for further evaluations on those tools to obtain additional results.

\paragraph{There is no one single solution in programming} 
One programming problem can be solved by many solutions. In addition, when considering the quality of a software or the quality of a source code, the result depends on subjective evaluation. 
For example, if maintainability is one of the key quality attributes needed for a software project, it is important to be clear on what quality metrics should be measured and the decision criteria to be used. 

The same idea is applied to grading student works. 
Each instructor defines grading criteria for an assignment. 
This research defines criteria to evaluate \emph{correct}, \emph{incomplete}, and \emph{incorrect} results. 
When the definitions and the criteria are different, the evaluation results also indicate otherwise. 
There are still subjectivity issues involved. Therefore, the experiment does not compare the quality of the generated code, refactored code, and test code with the results from other tools. 
The experiments analyzed the generated code and measured a wide range of software metrics to demonstrate that the responses from \chatgpt{} is not an out-of-the-box solution.  

\paragraph{The formulated series of questions might not cover all of the use cases}
There are other areas of competencies in computing-related education that are not included in this research. 
There is a possibility that \chatgpt{} lends itself well in other scenarios and use cases, but they are not tested by this research. 
To cover all of the competency areas, more experiments are required; which can be considered as a research opportunity for the computing research community.

\section{Advanced Topics in Computer Science}\label{sec:quantum}
Advanced topics in computer Science are somewhat uneasy to define.
In many cases, they require the use of advanced theoretical notations and specialized knowledge. 
For example, formal methods, static analysis, software architecture, cryptography, symbolic artificial intelligence, parallel and distributed computing, quantum informatics all contain some parts with much theoretical background.
All these courses are part of the curriculum of the Masters in Computer Science, Software Engineering, and Leadership from Constructor Institute.

In this study, we picked the topic that is the most difficult to tackle for our master students: quantum computing. Using principles from computer science, physics, and mathematics, quantum computing is an interdisciplinary domain that harnesses quantum mechanics to efficiently solve intricate problems.
Several software tools have already been developed for quantum computing, such as Qiskit \citep{Qiskit} by IBM Research, Silq \citep{SILQ} by ETH Zürich, or Cirq \citep{CIRQ} by by Google AI Quantum Team. 
This section examines the potential of \chatgpt{} as a tool for supporting research and education in quantum computing, an advanced topic in computer science.

\subsection{Experiment} \label{subsec:quantum_experiment}
To evaluate if \chatgpt{} is reliable when asked about quantum computing, the experimenter asks simple questions about quantum gates and analyses the answers given by \chatgpt{}. 
The gates considered consist of the identity gate, all one-qubit gates, and all two-qubit gates mentioned as common by Wikipedia \citep{WikipediaQuantumGate}. 
For each of these quantum gates, the experimenter asks \chatgpt{} three elements:
\begin{enumerate}
    \item a \emph{definition} of the gate, 
    \item a \emph{circuit} representation, and
    \item the \emph{final state} after applying the quantum gate to a given initial state.
\end{enumerate}
The considered initial states are all states of the computational basis of the corresponding dimension as well as an arbitrary state with free parameters.

The three elements asked for each gate --- definition, circuit, final state --- allow the evaluation of \chatgpt{} in three different aspects. 
Because the definition is text-based, the circuit is a concept visualization, and the computation of a state relies on logic. 
The three elements together provide a complete understanding of the quantum gate.

Answers are evaluated based on the different elements independently and classified as \textit{correct}, \textit{incomplete} or \textit{incorrect} as described in Section \ref{sec:methodology}. \newline

\begin{figure*}[tb!]
  \centering
  \includegraphics[width=0.7\textwidth]{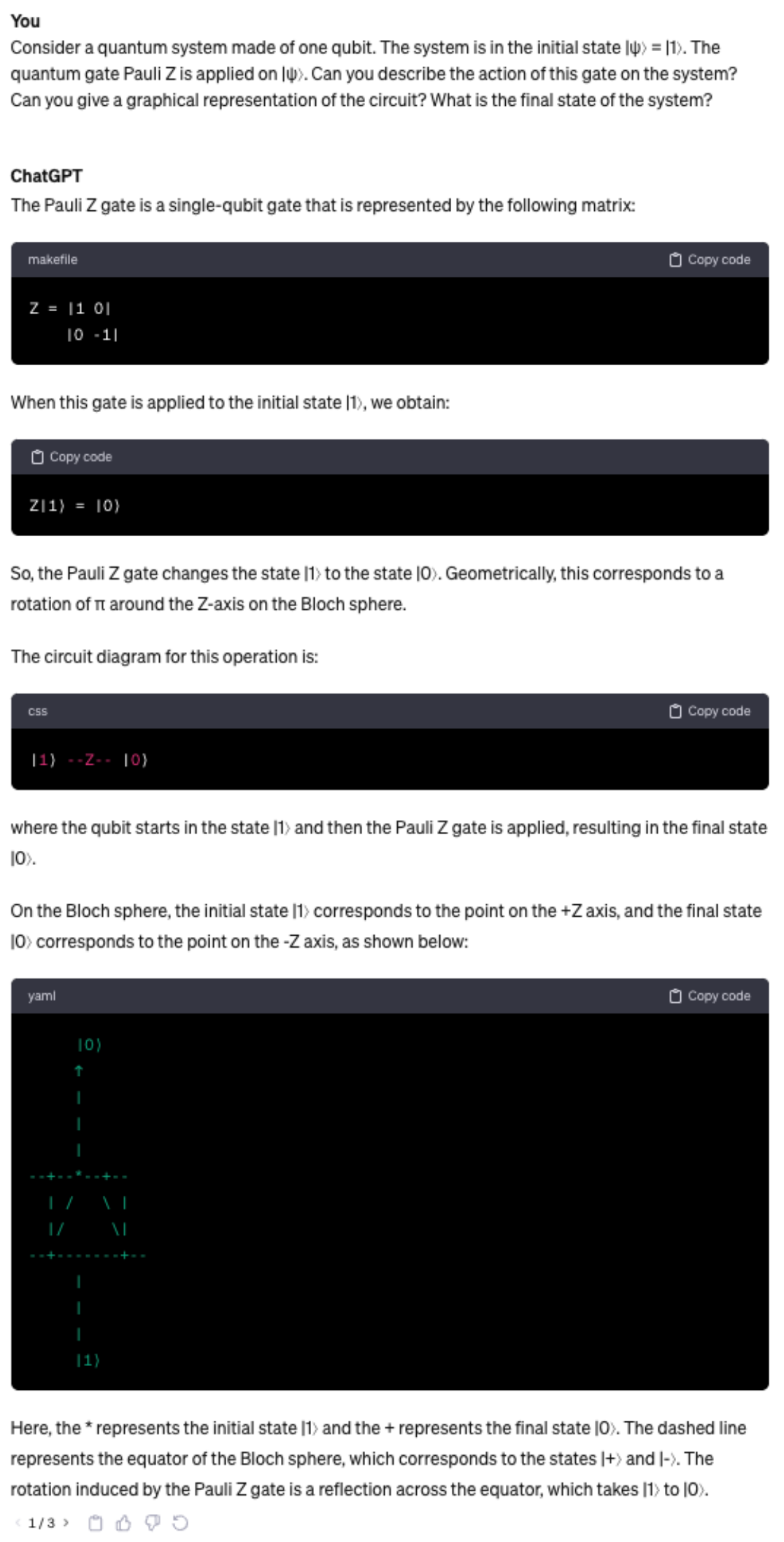}
\caption{First answer of \chatgpt{} to questions about the Pauli X with initial state $\ket{1}$. }
\label{f:pauli_Z_1_1}       
\end{figure*}

\begin{figure*}[tb!]
  \centering
  \includegraphics[width=0.7\textwidth]{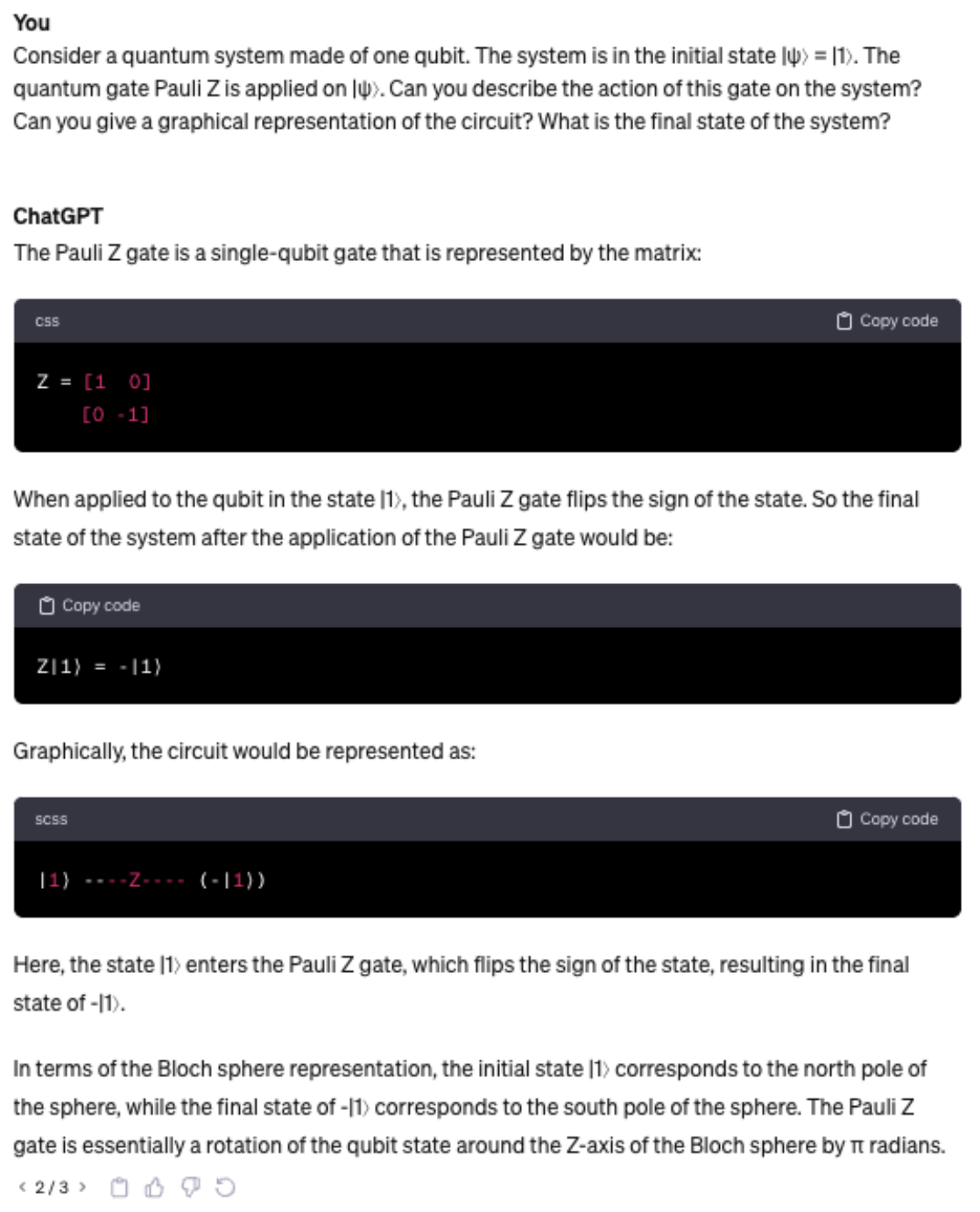}
\caption{Second answer of \chatgpt{} to the question about the Pauli Z with initial state $\ket{1}$. }
\label{f:pauli_Z_1_2}       
\end{figure*}

\begin{figure*}[tb!]
  \centering
  \includegraphics[width=0.7\textwidth]{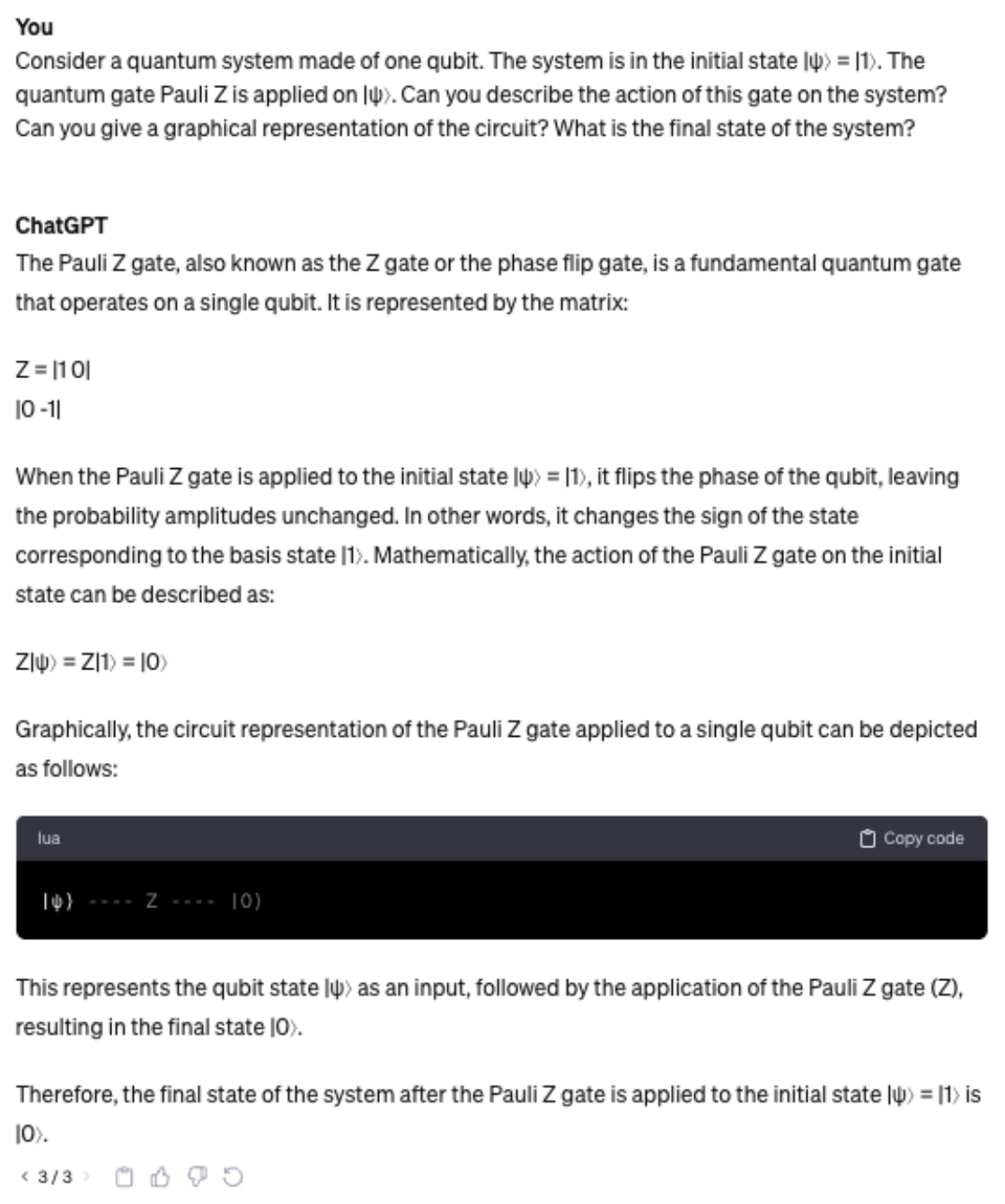}
\caption{Third answer of \chatgpt{} to the question about the Pauli Z with initial state $\ket{1}$. }
\label{f:pauli_Z_1_3}       
\end{figure*}

Figures \ref{f:pauli_Z_1_1}, \ref{f:pauli_Z_1_2} and  \ref{f:pauli_Z_1_3} serve as an example of the methodology. 
The question is about the gate Pauli Z and the initial state is $\ket{1}$, so the question asked to \chatgpt{} is:

\emph{Consider a quantum system made of one qubit. The system is in the initial state $\ket{\psi}=\ket{1}$. The quantum gate Pauli Z is applied on $\ket{\psi}$. Can you describe the action of this gate on the system? Can you give a graphical representation of the circuit? What is the final state of the system?}

We can see in Fig.~\ref{f:pauli_Z_1_1} that the three question elements are incorrect since $\chatgpt{}$ fails to describe the effect of Pauli Z on $\ket{1}$. 
The second attempt of \chatgpt{} (Figure \ref{f:pauli_Z_1_2}) yields correct answers for both the circuit and the final state. 
The definition, however, must be considered incomplete. 
Despite the correct description of the effect of Pauli Z, it includes extra information about the Bloch sphere, which are erroneous.
Figure \ref{f:pauli_Z_1_3} illustrates the third response generated by \chatgpt{}. 
In this instance, the definition as well as the circuit of the Pauli Z gate are correct. The final quantum state is, however, incorrect. 
Note that, in cases where the graphical representation of the gate is precise, the circuit is deemed correct, irrespective of any inaccuracies in the displayed quantum states.

\subsection{Results} \label{subsec: quantum_results}
As described in Section \ref{sec:methodology}, the experiment computes the percentages of correct answers, incomplete answers and an incorrect answers. 
In practice, the experiment considers independently each question element --- definition, circuit, final state. 
Table \ref{t:quantum_probs} shows the results.

\begin{landscape}
\begin{table}
    \centering
    \caption{Percentages of correct answers, incomplete answers or incorrect answers when asking \chatgpt{} the definition, circuit and application of each quantum gate. It also includes the global average and the averages over one-qubit gates and two-qubit gates independently. For control gates, the first number denotes the control qubit, while the second denotes the target qubit.}
    \label{t:quantum_probs}
    \begin{tabular}{cccccccccc}
    \toprule
                             & \multicolumn{3}{c}{Definition}            & \multicolumn{3}{c}{Circuit}               & \multicolumn{3}{c}{Final state}           \\ \hline
    \multirow{2}{*}{Observation} & \multicolumn{3}{c}{Percentages} & \multicolumn{3}{c}{Percentages} & \multicolumn{3}{c}{Percentages} \\ \cmidrule(rr){2-4} \cmidrule(rr){5-7} \cmidrule(rr){8-10} 
                                 & Correct  & Incomplete  & Incorrect  & Correct  & Incomplete  & Incorrect  & Correct  & Incomplete  & Incorrect  \\ \hline
    Global                       & 84\%       & 3\%          & 13\%         & 37\%       & 20\%          & 43\%         & 70\%       & 5\%          & 25\%         \\ \hline
    1-qubit gates                & 89\%       & 5\%          & 6\%         & 67\%       & 22\%          & 11\%         & 74\%       & 2\%          & 24\%         \\
    2-qubit gates                & 80\%       & 1\%          & 19\%         & 12\%       & 17\%          & 71\%         & 67\%       & 7\%          & 26\%         \\ \hline
    Identity                     & 100\%       & 0\%          & 0\%         & 78\%       & 22\%          & 0\%         & 89\%       & 11\%          & 0\%         \\
    Pauli X                      & 100\%       & 0\%          & 0\%         & 78\%       & 11\%          & 11\%         & 100\%       & 0\%          & 0\%         \\
    Pauli Y                      & 100\%       & 0\%          & 0\%         & 67\%       & 22\%          & 11\%         & 67\%       & 0\%          & 33\%         \\
    Pauli Z                      & 67\%       & 11\%          & 22\%         & 67\%       & 11\%          & 22\%         & 56\%       & 0\%          & 44\%         \\
    Hadamard                     & 78\%       & 11\%          & 11\%         & 78\%       & 22\%          & 0\%         & 67\%       & 0\%          & 33\%         \\
    Phase gate                   & 89\%       & 0\%          & 11\%         & 34\%       & 33\%          & 33\%         & 78\%       & 0\%          & 22\%         \\
    T gate                       & 89\%       & 11\%          & 0\%         & 67\%       & 33\%          & 0\%         & 67\%       & 0\%          & 33\%         \\
    CNOT$(1, 2)$                 & 80\%       & 7\%          & 13\%         & 13\%       & 20\%          & 67\%         & 93\%       & 0\%          & 7\%         \\
    CNOT$(2, 1)$                 & 67\%       & 0\%          & 33\%         & 0\%       & 33\%          & 67\%         & 40\%       & 7\%          & 53\%         \\
    CZ$(1,2)$                    & 67\%       & 0\%          & 33\%         & 20\%       & 7\%          & 73\%         & 60\%       & 7\%          & 33\%         \\
    CZ$(2,1)$                    & 93\%       & 0\%          & 7\%         & 27\%       & 20\%          & 53\%         & 60\%       & 7\%          & 33\%         \\
    Swap                         & 93\%       & 0\%          & 7\%         & 0\%       & 7\%          & 93\%         & 80\%       & 13\%          & 7\%         \\    
    \bottomrule
    \end{tabular}
\end{table}
\end{landscape}

Global estimations show that \chatgpt{} gives a correct answer more often when it is asked for the definition of the quantum gate than for the circuit or the final state in 84\% of the cases. 
This is an expected result since, as mentioned before, a definition is text-based and \chatgpt{} is a natural language processing tool.
The most common mistake for circuits of one-qubit gate was to omit the final state.
For two-qubit-gate circuit no specific error pattern could be detected.

When comparing the percentages of correct answers of one-qubit gates and two-qubit gates, one can see that the former are at least 8\% larger than the latter. 
This difference is especially high for the circuit element.

Upon closer examination of the outcomes concerning one-qubit gates, it becomes evident that the Pauli X gate exhibits the highest percentages of correct answers for all three components. The results obtained from the Pauli Y gate are similar to those of the Pauli X gate, but there is a significant decrease in the percentages of correct answers for the Pauli Z gate. 
This decrease can be primarily attributed to ChatGPT applying a Pauli X gate instead of a Pauli Z gate.

When applying control gates, the roles of each qubit is important as in general C$U$($q$,$p$ $\neq$ C$U$($p$,$q$), where $U$ refer to a one-qubit unitary and the control and the target qubit are assumed to be the first and the second qubit, respectively. 

Surprisingly, in the case of the CNOT gate, the responses on CNOT(1,2) are, in general, more often correct than the answers on CNOT(2,1)! 
When \chatgpt{} gave an incorrect final state for CNOT(2,1), the mistake was exchanging the control and the target qubit. 
In the literature, it is more typical to apply the CNOT considering the first qubit as the control qubit.
This could explain the discrepancy of results between CNOT(1,2) and CNOT(2,1).

For the control phase gate, CZ, it holds that $\text{CZ}(1,2)\ket{\psi} = \text{CZ}(2,1)\ket{\psi}$ for any state $\ket{\psi}$. 
It is not possible to determine whether \chatgpt{} has swapped the control and target qubits looking only at the final state, but we need explicit statements. 
Examining the percentages for the definition and the circuit of CZ in Table \ref{t:quantum_probs}, we observe that CZ(2,1) exhibits higher percentages of correct answers for both the definition and circuit, yet the percentages for the final state remain unchanged. 
The primary mistake in these instances is confusing a phase shift with a phase factor.

As a conclusion, \chatgpt{} shows relatively good results when producing clear text explanations of the gates.
For the rest of the answers, it generates poor-quality results.

\subsection{Limitations and threats to validity} \label{subsec:threats}
In the evaluation of \chatgpt{} as a support tool for research and education in quantum computing, \chatgpt{} returns more often a correct answer when asked for the definition of a quantum gate than for the circuit or the final state. The percentages of correct answers for the definition is only 84\%.
This means that \chatgpt{} cannot be a main tool for research and education in quantum computing.
At best, we can only imagine it as a test generation tool to find potentially incorrect answers for students to fix.

Several questions remain open. Would \chatgpt{} become a reliable support tool for teaching and research in quantum computing if it had access to logic? In March 2023, a plugin for \chatgpt{} that connects it with WolframAlpha was released~\citep{WolframAlphaPlugin}. 
This plugin provides \chatgpt{}4 access to the well-known Wolfram Language, and promises that using this plugin \chatgpt{}4 can do nontrivial computations as well as systematically produce correct data. 
Since all computations evaluated \chatgpt{}3.5 can be done using matrix multiplications, it would be promising to repeat our evaluation using the WolframAlpha plugin. 
A preliminary research shows that ChatGPT4 gives more often a correct answer than ChatGPT3.5 for one-qubit quantum gates, while results seem similar for two-qubit quantum gates and concatenations of gates. 
Further research should be performed to confirm it.

This study uses the quantum computing model based on quantum gates, but there exists others such as one-way quantum computers or adiabatic quantum computers. 
Would ChatGPT perform better if we considered another model of quantum computing? 
Further research should also be performed to answer that question.

\section{\review{Implications for Education in Computer Science}}\label{sec:implications}

\review{
After working extensively on these topics, we believe that our work has several implications for the near future:
}
\begin {itemize}
\item \review{Regarding the quality of student education: simple questions can be answered automatically by ChatGPT and, by extension, other conversational agents. Such agents can, already now, be considered as a good way to help first year students self-check and get started. The use of such tools, could then be encouraged as they give an answer faster than a teaching assistant. The discussion whether it is better for students to look for answers on their own is, in our opinion, sterile as Web search engines already eased much access to the information over spending time reading books to find answers.}
\item \review{Regarding limitations of the tools: students who are trusting such solutions (as they use them on a regular basis) need to be warned about the dangers and trained in discovering their limitations. It is the duty of educators to help students understand the use and limitations of tools.}
\item  \review{Regarding assignments and assessments: many instructors chose to run assignments outside of the classroom. To encourage students to be critical and further their understanding, assignments need to change and be adapted to face the new reality: courses with repetitive exercises asking students to code simple data structures need to either run entirely without the help of the Internet altogether or be adapted to cope with the new reality to make sure that the tools cannot provide a quick and easy answer.}
\end{itemize}

\review{The following subsections present a tentative teaching sequence based on the experiments presented in this article showing the limitations of conversational agents as well as an example of assignment that makes students further their understanding of the programming exercise.} 

\subsection{\review{Teaching Sequence to find Limitations of Conversational Agents}}
\review{This section presents the teaching sequence that allows students to find limitations of conversational agents.}

\review{The exercices are presented below:}

\review{\textit{These exercises aim at making students aware of the strength and weaknesses of conversational agents as helpers for programming tasks}}

\review{\textit{Question 1: Use a conversational agent to obtain an implementation of a binary search tree in Java.}}

\review{Expected answer: a program that someone can copy/paste and arrange in the proper files to compile. The program will work.}

\review{\textit{Question 2: Can you compile the result?}}

\review{For students who do not have the JDK installed, time will be spent installing and finding out how to install components.
The program will be able to compile and run.}

\review{\textit{Question 3: Use a conversational agent to create test cases with a unit test framework for the binary search tree.}}

\review{Expected answer: Test cases will be created.}

\review{\textit{Question 4: Can you compile the results? In case you have issues use the conversational agent to obtain instructions.}}

\review{Once more, installing the necessary components might be tricky.
In our experience, this is a critical step on specific machines.
It is actually harder to use some packages and the conversational agent does not know them.}

\review{\textit{Question 5: Download an IDE and look up on a search engine for an implementation of a binary search tree, create a project with it and look up how to write a test case in the IDE. Was it easier to do that with the conversational engine or with the IDE and the search engine?}}

\review{In our experience, when using a given IDE, the documentation is much more explicit and resources are easier to find than by using the conversational agent.}

\review{\textit{Question 6: Use a conversational agent to create a mediator pattern? Can you modify it to make it thread-safe? Can you understand what the singleton pattern is doing and can you identify issues with the design of the code? Use a tool to assess the quality of the code.}}

\review{As mentioned in this article, many issues exist with the maintainability of the resulting code.}

\review{\textit{Question 7: Use a conversational agent to create quantum code that applies a Hadamard gate to the state $\ket{1}$. Ask to the conversational agent "are you sure?". Is the result the same? Can you find out which one is correct?}}

\review{As mentioned in this article, many issues exist with the results for quantum calculations. Suprisingly the results given by the conversational agent differ after simply asking "asks are you sure?". The real question is whether we can trust the answers at all if they differ and we have no idea of the proper answer.}

\review{\textit{Question 8: Search on the Internet for "ChatGPT fails". Can you reproduce five fails? After careful evaluations, discuss in groups of 4 what you think about the usefulness of conversational agents and their limitations? Can you devise a strategy to avoid mistakes when using conversational agents?}}

\review{Some of the failures are puzzling because of their simplicity, reproducing them is easy and proving that you cannot trust blindly the results returned by ChatGPT. Students should arrive to this conclusions and find strategies to avoid the pitfall.}

\review{We plan to run such an evaluation as an activity with our students with every new cohort.}

\subsection{\review{Example of Assignments that Further the Students' Understanding}}

\review{In this section we ask the student to use usual mistakes by students to fool ChatGPT. We believe that this experience can enhance the students' understanding about the limitations of ChatGPT. A possible exercise could be the following.}

\review{\textit{Example: What is the most common mistake that programmers do when trying to exchange variables? Can you make ChatGPT generate the wrong explanation? Explain why the answer generated by ChatGPT is wrong.}}

\review{Generating such examples is surprisingly easy. Here are a few statements that can be falsified easily by ChatGPT:}

\begin{figure*}[tb!]
  \centering
  \includegraphics[width=0.7\textwidth]{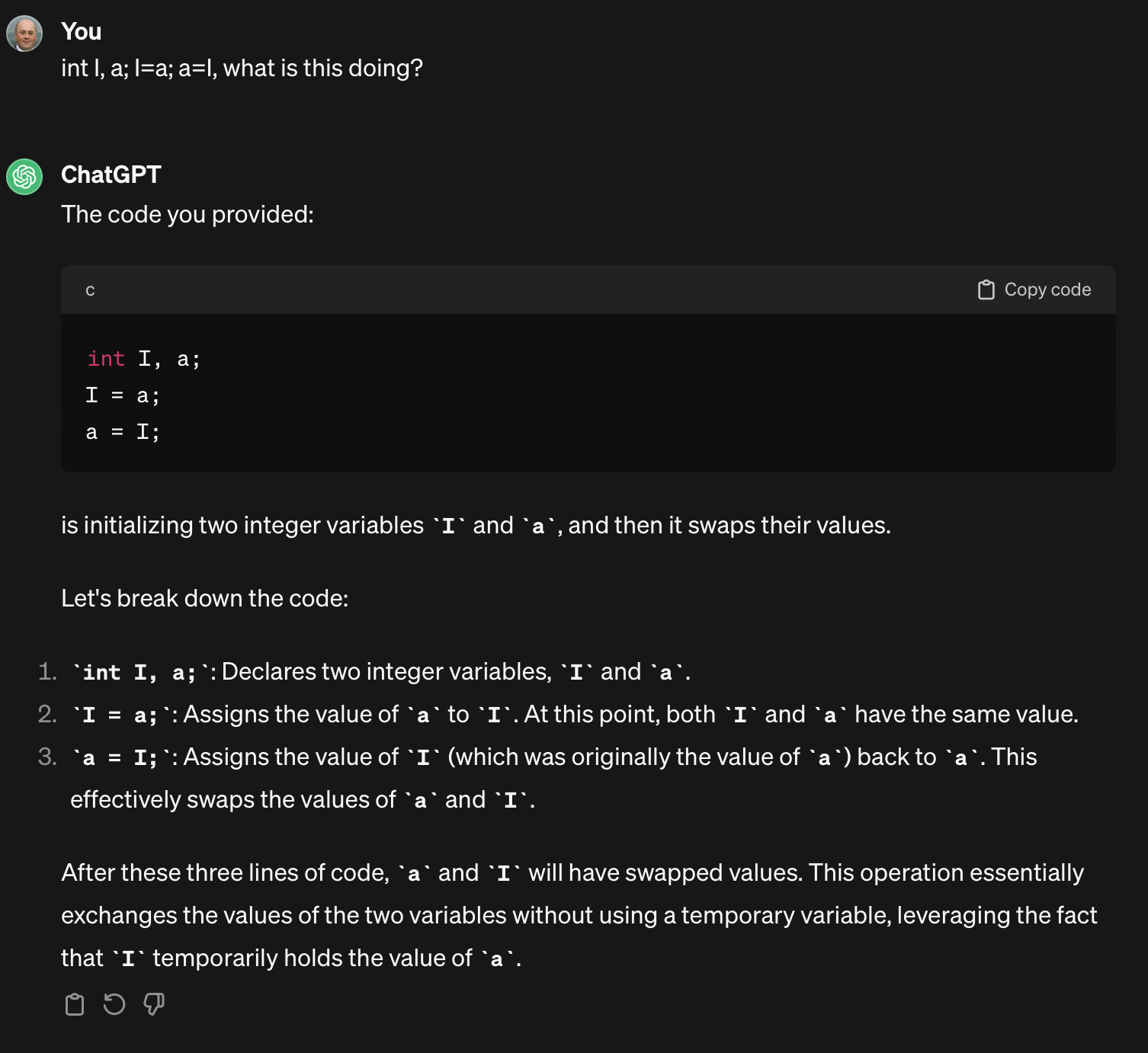}
\caption{ChatGPT generating a wrong interpretation about not swapping variables. }
\label{f:swap}       
\end{figure*}

\begin{itemize}
    \item \review{What is this doing: int a, b; a=b, b=a? (answer is wrong because of assignment erasing the other value, see Fig.~\ref{f:swap})}
    \item \review{int a; is a*a positive? (answer is wrong because of overflows)}
    \item \review{int a,b; is (a*b)/b equal to a? (answer is wrong because of overflows)}  
    \item \review{paste some code that shows the diamond problem in C++, and remove the main function. Ask if this compiles. (ChatGPT says no, but if there is no use of the code, it compiles)}
\end{itemize}

\review{Students trying to fault ChatGPT will learn more about its limitations, and the finesse of the problem at hand than if the questions are simple to answer.}

\section{Related Work}\label{sec:related_work}

Bots are becoming available to software engineers willing to improve their productivity~\citep{8823643,mapping}.
For example, \citet{DBLP:journals/tse/CarrLP17} created a bot that inserts automatically proven contracts in source code, \citet{8115628} made a chat bot that answers questions about APIs, \citet{10.1145/3180155.3180238} made a development assistant able to understand commands for Git and GitHub tasks. 
\review{\cite{10.1145/3511861.3511863} evaluated the results from Codex in programming CS1 tests and even solving a Rainfall Problem.}
\review{In 2021, GitHub Copilot was introduced as an AI programming assistant that can suggest code completion.} \review{\citet{10.1145/3491101.3519665} showed that despite the average results in terms of time and correctness of the generated code, it is at least a good starting point to approach the programming task.} \review{However, \citet{liang2024} demonstrated that developers utilize AI programming assistants primarily to minimize keystrokes, expedite programming tasks, and retrieve syntax. Their research indicates less enthusiasm for employing these assistants in brainstorming potential solutions. This reluctance stems from the fact that these tools often fail to produce code that adequately addresses specific functional or non-functional requirements, and developers encounter challenges in controlling the tool to generate desired outputs.} 

For automated bots generating code, most articles tend to focus on making it as close to what a programmer could have generated. For example, generating automatically patches as well as explanations to help developers understand them~\citep{8823632,10.1145/3183519.3183540} or make refactorings indistinguishable from what a human could have generated~\citep{8823629}. \review{Owura at al. (2023) showed that Copilot is less likely to generate the same vulnerabilities as human developers, implying that Copilot is not always as bad as human software developers. }

The computer science education community has been mostly unaware of the potential of \genai{} until 2023. 
Since the beginning of 2023, many see in \chatgpt{}, and generally in AI, a transforming technology for teaching and research~\citep{murugesan2023, damian2023, yilmaz2023, lim2023}.
We performed an exhaustive search on articles for the past 20 years in the proceedings of the two premium computer science education conferences: the Symposium of the ACM Special Interest Group in Computer Science Education (SIGCSE)\footnote{https://www.sigcse.org} and the conference in Innovation and Technology in Computer Science Education (ITICSE)\footnote{https://iticse.acm.org}. 
This search (using the keyword ``chat'' in the title to pre-filter articles) returned exactly four research articles~\citep{10.1145/3545947.3576285,10.1145/1352135.1352287,10.1145/3341525.3393979,10.1145/2591708.2591728} and two abstract~\citep{10.1145/2839509.2850526,10.1145/3408877.3439693} mentioning the use of chatbots in CS education.

In these articles, three articles mention the boost on motivation or well-being of students~\citep{10.1145/3545947.3576285,10.1145/1352135.1352287,10.1145/2591708.2591728} by using or specializing chatbots (similarly to \citet{yilmaz2023}). 
One of the abstracts mentions the use of chatbots for teaching inclusion~\citep{10.1145/3408877.3439693} and the last one is used to automate tasks in the classroom that are either tedious or error-prone~\citep{10.1145/2839509.2850526}. 
Such applications, while useful, are not directly related to teaching programming using chatbots.
The last article~\citep{10.1145/3341525.3393979} in the series presents a chatbot that asks questions to the users to establish a profile used to teach specific skills.

\review{\cite{Laato2023} showed that AI can be used as learning by asking questions(as kids usually do) and, also contains an enormous library of knowledge, nothing equally powerful has been made available to the general public in this scale before. \cite{10132255} evaluated the quality of the ChatGPT answering questions
involving both code and concepts. 
\cite{Wang2023} stated that students can have more personalized learning experience with LLMs, but the curriculum should be adopted with AI-driven tools and strategies. Also \cite{Tian2023} discovered that ChatGPT is effective in dealing with common programming problems and identified the ability of ChatGPT to reason the original intention of the code, but there are limitations in terms of its attention span, that detailed descriptions can constrain the focus of ChatGPT and prevent it from leveraging its vast knowledge to solve the actual problem.
\cite{Richards2023} also proved that in most cases, across a range of question formats, topics, and study levels, ChatGPT is at least capable of producing adequate answers for undergraduate assessment. Moreover, \cite{Malinka2023} showed that AI might pass the courses required for a university degree and this can threaten the traditional way of evaluating students. The positive impact is that an AI assistant can help to accelerate the learning process and even supplement the teacher.}

This list will undoubtedly grow to include many articles on the use of chatbots in education published in 2023 and later.\footnote{Article originally submitted in July 2023.}
For example, the list of sessions in the coming edition of ITiCSE,\footnote{https://iticse.acm.org/2023/conference-program/} whose program is already announced (but not yet published) contains already 3 articles and one session on the use of a chatbot in CS education. 

Along with some of the authors' other publication~\citep{fisee23Kotovich}, which evaluates whether students can generate fundamental algorithms using ChatGPT3 and whether they can be caught for plagiarism, the present article is one of the first publications evaluating the usefulness of ChatGPT for teaching with clear pros and cons. We believe it will not be the last of its kind.  \review{There are ethical implications and concerns regarding the integration of AI in undergraduate computer science education. However, positive outcomes can be achieved with proper planning, guidelines, and support~\citep{yunkailiu2023, 10305701}. LLM detectors like \citet{copyleaks}, \citet{gptkit}, \citet{gptzero}, and GLTR~\citep{gehrmann-etal-2019-gltr, gehrmann2019gltr} are introduced to relieve the threats of \genai{} to academic integrity. There are cases that the LLM detectors work and those that do not.}

\review{In recent years, there has been an increasing amount of research focusing on the incorporation of AI into computer science education~\citep{10213396, 10305701, 10260931, 10342970, yunkailiu2023}. Research has delved into different facets of AI within education, covering areas such as the influence of AI technologies on the curriculum, the efficacy of pedagogical methods based on AI, and the hurdles encountered in integrating AI into computer science education. \citet{10342970} conducted an initial investigation into the efficacy of \chatgpt{} in improving undergraduate computer science and engineering education. The students involved in the study suggested that ChatGPT's usefulness is limited or entirely negligible due to its unreliable results. Unlike others, we demonstrated the benefits and risks of using \chatgpt{} as a support tool for teaching computer science through our experiments and evaluations on the effectiveness of \chatgpt{} in solving computer science problems at three different levels.}

\section{Conclusions}\label{sec:conc}

Some tools suddenly open possibilities that we thought would never be reality.
\chatgpt{} is one of these tools.
It suddenly sparked a very strong interest and captured the imagination of many.
Is it interesting? Yes! 
Is it the silver bullet that it appears to be at first? No yet!

The present article evaluates the performance of \chatgpt{} as a support tool for teaching topics related to the computer science curriculum.
It evaluates the answers of \chatgpt{} at three levels: fundamental (first-year students), core (finishing bachelor students), and advanced (master-level students).
For fundamental knowledge, the evaluation focuses on answers about standard algorithms and data structures.
For core knowledge, it explores the answers on design patterns.
For advanced knowledge, it studies answers on one and two qubits gates in quantum computing.

The main result of this study is that the output of \chatgpt{} contains more errors as the questions target more advanced topics: it answers fundamental questions almost perfectly, it answers questions about design patterns mostly correctly even if it contains code smells and generally poor quality code, it has generally poor performances on quantum computing.
The most concerning in this is that it always answers questions with the highest level of certainty, even when it is completely wrong. 

We have no doubts that \chatgpt{} will improve in the future (and we will undoubtedly reevaluate further tools and further versions of the tools).
For the time being, however, we believe that students can use \chatgpt{} for their studies, but they need to be warned.
Even if, at first, they can use it for basic requests, as they progress in their curriculum, they will need to question its answers because they will contain more and more incorrect statements.

\appendix

\bibliographystyle{elsarticle-harv} 
\bibliography{biblio}

\section{Quantum gates}
\label{appendix:quantum_gates}
In Section \ref{sec:quantum}, we have tested \chatgpt{} capabilities concerning quantum computing. In concrete, we have considered common quantum gates and asked for their definition, circuit representation and application on a given quantum state. This section contains the theoretical background behind the questions.\newline

Below, we use the computational basis denoted by $\{\ket{0}, \ket{1}\}$, and corresponding tensor products for higher dimensions. We refer to one- and two-qubit arbitrary quantum states as $\ket{\psi}$, which can be expanded in the corresponding computational basis as
\begin{equation} \label{e:one-qb_state}
    \ket{\psi} = \dfrac{1}{\sqrt{N}}(a\ket{0}+b\ket{1}),    
\end{equation}
\begin{equation} \label{e:two-qb_state}
    \ket{\psi} = \dfrac{1}{\sqrt{N}}(a\ket{00}+b\ket{01}+c\ket{10}+d\ket{11}),
\end{equation}
respectively. Here, $a, b, c, d$ are complex numbers, and $N$ is the normalization factor satisfying that $|\braket{\psi}|$.
\newline



The most simple, but essential quantum gate is the so-called Identity. The Identity gate, $I$,  is a one-qubit gate that does not affect the input state. Mathematically,
$$
I\ket{\psi} = \ket{\psi} \forall \ket{\psi}.
$$
The circuit representation of the Identity gate, as well as the upcoming gates, can be found in Table~\ref{t:circuits}.

The Pauli X, $X$, also known as bit-flip gate, is a one-qubit gate that exchanges the states of the computational basis. In other words,
$$
X\ket{0} = \ket{1},
$$
$$
X\ket{1} = \ket{0}.
$$
Therefore, given an arbitrary state in the computational basis, we have

$$
X \ket{\psi} = \dfrac{1}{\sqrt{N}}(a\ket{1} + b\ket{0}).
$$

The Pauli Y, $Y$, is a one-qubit gate that exchanges the states of the computational basis while introducing a phase factor as
$$
Y\ket{0} = i\ket{1},
$$
$$
Y\ket{1} = -i\ket{0}.
$$
Given an arbitrary state in the computational basis, the resulting state after applying a Pauli Y is
$$
Y \ket{\psi} = \dfrac{i}{\sqrt{N}}(a\ket{1} - b\ket{0}),
$$
where the global phase factor $i$ can be absorbed by the normalization factor and it has no physical meaning. Note that the relative phase factor $-1$ does have implications in the physical properties of $\ket{\psi}$.

The last Pauli gate is the Pauli Z, $Z$, which is also referred to as phase flip. The action of Pauli Z on the computational basis is
$$
Z\ket{0} = \ket{0},
$$
$$
Z\ket{1} = -\ket{1}.
$$
In other words, the $Z$ gate leaves $\ket{0}$ unchanged and flips the phase of $\ket{1}$. Thus, the Pauli Z transforms an arbitrary state to

$$
Z \ket{\psi} = \dfrac{1}{\sqrt{N}}(a\ket{0} - b\ket{1}).
$$

The Hadamard gate, $H$, is a one-qubit gate that takes the computational-basis states and creates an equal-superposition state as

$$
H\ket{0} = \dfrac{1}{\sqrt{2}}(\ket{0}+\ket{1}),
$$
$$
H\ket{1} = \dfrac{1}{\sqrt{2}}(\ket{0}-\ket{1}).
$$
Given an arbitrary state $\ket{\psi}$, the application of a Hadamard gate yields to

$$
H \ket{\psi} = \dfrac{1}{\sqrt{2N}}\left[(a+b)\ket{0} + (a - b)\ket{1}\right].
$$

The Phase shift is a family of one-qubit gates denoted by $P(\phi)$. It introduces a relative phase factor of $\phi$ between the computational-basis states. Mathematically,

$$
P\ket{0} = \ket{0},
$$
$$
P\ket{1} = e^{i\phi}\ket{1}.
$$
The action of a phase shift on an arbitrary quantum state is
$$
P \ket{\psi} = \dfrac{1}{\sqrt{N}}\left(a\ket{0} + be^{i\phi}\ket{1}\right).
$$

For some values of the phase factor $\phi$, the phase shift has a concrete name. For example, $P$ is known as the phase gate or the S gate, $S$, when $\phi = \pi/2$. The phase shift $P(\pi/4)$ is referred to as the T gate or the $\pi/8$-gate. Note that $Z = P(\pi)$, $S = P(\pi/2) = \sqrt{Z}$ and $T=P(\phi/4)=\sqrt[4]{Z}$.

The controlled NOT gate, CNOT$(c,t)$, is a two-qubit quantum gate that applies a bit flip to the target qubit, $t$, if and only if the control qubit, $c$, is in $\ket{1}$. Otherwise, the target qubit is unchanged. Considering the first qubit as the control qubit and the second qubit as the target qubit, the application of the CNOT on the computational basis results in
$$
\text{CNOT}(1,2)\ket{00} = \ket{00},
$$
$$
\text{CNOT}(1,2)\ket{01} = \ket{01},
$$
$$
\text{CNOT}(1,2)\ket{10} = \ket{11},
$$
$$
\text{CNOT}(1,2)\ket{11} = \ket{10}.
$$
If we apply the CNOT gate on an arbitrary two-qubit state (Eq. (\ref{e:two-qb_state})), we obtain
$$
\text{CNOT} \ket{\psi} = \dfrac{1}{\sqrt{N}}\left(a\ket{00} + b\ket{01} + c\ket{11} + d\ket{10}\right).
$$

Analogously to the controlled NOT gate, there exists the controlled phase gate, CZ$(c,t)$, which applies a phase flip to the target qubit, $t$, if and only if the control qubit, $c$, is in $\ket{1}$. Mathematically, we write
$$
\text{CZ}(c,t)\ket{00} = \ket{00},
$$
$$
\text{CZ}(c,t)\ket{01} = \ket{01},
$$
$$
\text{CZ}(c,t)\ket{10} = \ket{10},
$$
$$
\text{CZ}(c,t)\ket{11} = -\ket{11}.
$$
Note that the controlled phase gate is independent of the role of each qubit, i.e., CZ$(c,t)$ = CZ$(t,c)$. After applying the CZ gate on an arbitrary two-qubit state in Eq.~(\ref{e:two-qb_state}), we obtain
$$
\text{CZ} \ket{\psi} = \dfrac{1}{\sqrt{N}}\left(a\ket{00} + b\ket{01} + c\ket{01} - d\ket{11}\right).
$$

The SWAP gate is a two-qubit gate that swaps the state of the qubits on which it acts. Applying the SWAP gate on the computational basis yields to
$$
\text{SWAP}\ket{00} = \ket{00},
$$
$$
\text{SWAP}\ket{01} = \ket{10}.
$$
$$
\text{SWAP}\ket{10} = \ket{01}.
$$
$$
\text{SWAP}\ket{11} = \ket{11}.
$$
Therefore, if we apply the SWAP gate on an arbitrary two-qubit state, we get
$$
\text{SWAP} \ket{\psi} = \dfrac{1}{\sqrt{N}}\left(a\ket{00} + b\ket{10} + c\ket{01} + d\ket{11}\right).
$$

\begin{table}[]
\bgroup
\def\arraystretch{2}
\begin{tabular}{cc}
\textbf{Quantum gate}              & \textbf{Circuit representation(s)} \\
Identity ($I$)   &    \parbox[c]{7cm}{\centering\includegraphics[scale=0.075]{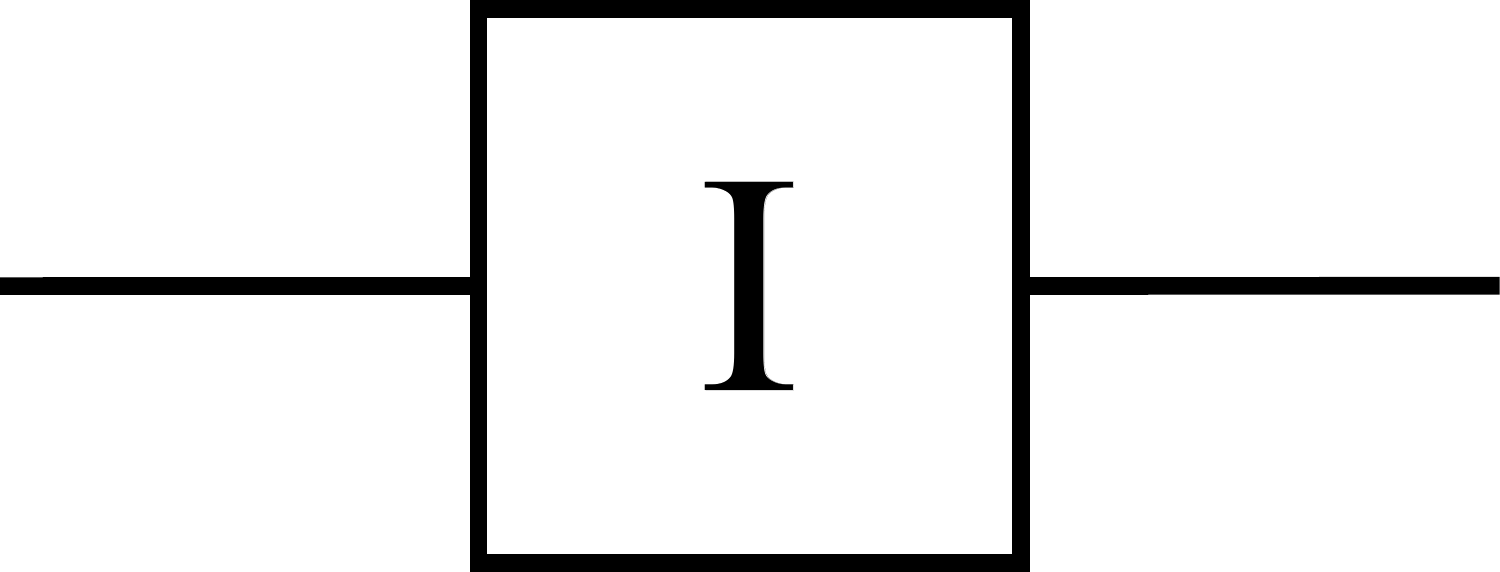}} \\
Pauli X ($X$)    &    \parbox[c]{7cm}{\centering\includegraphics[scale=0.075]{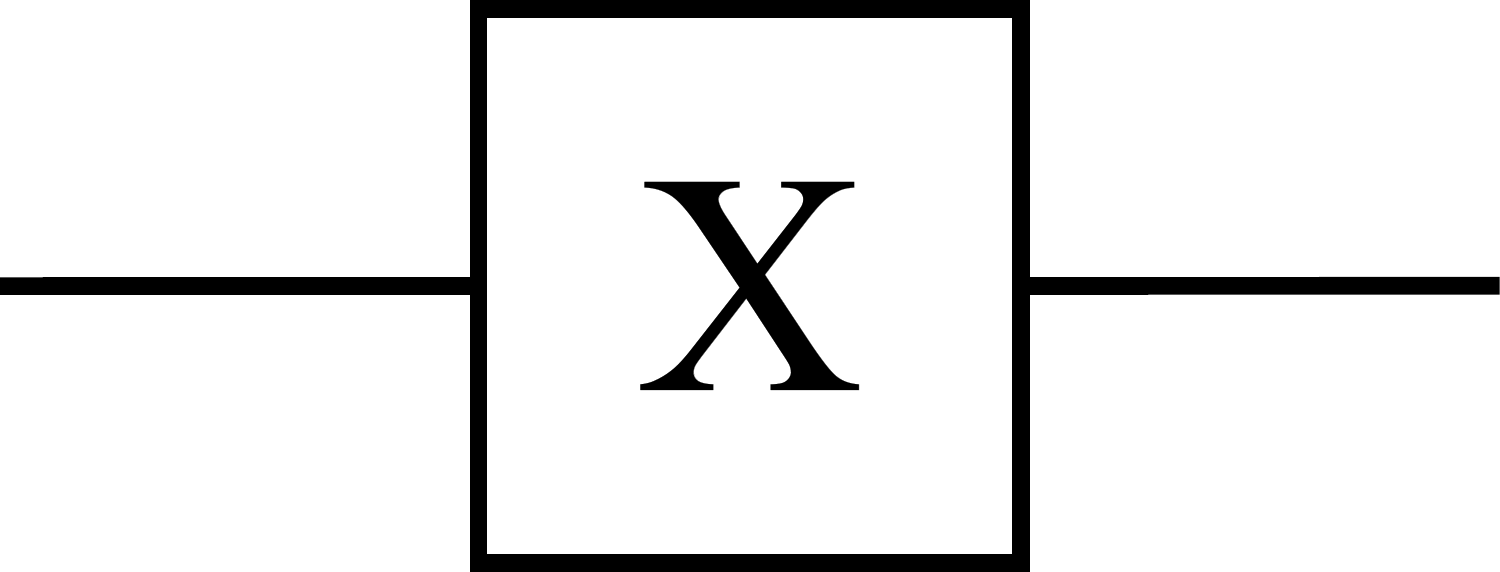} \hspace{0.2cm}\includegraphics[scale=0.075]{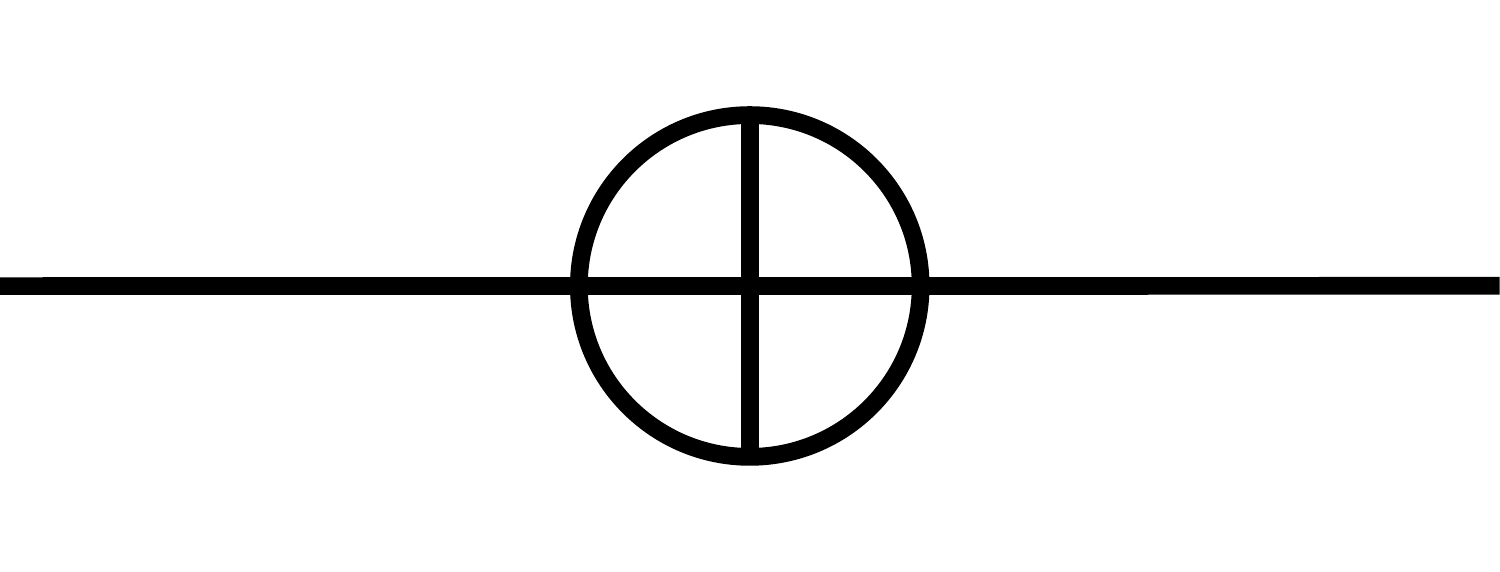}} \\
Pauli Y ($Y$)    &    \parbox[c]{7cm}{\centering\includegraphics[scale=0.075]{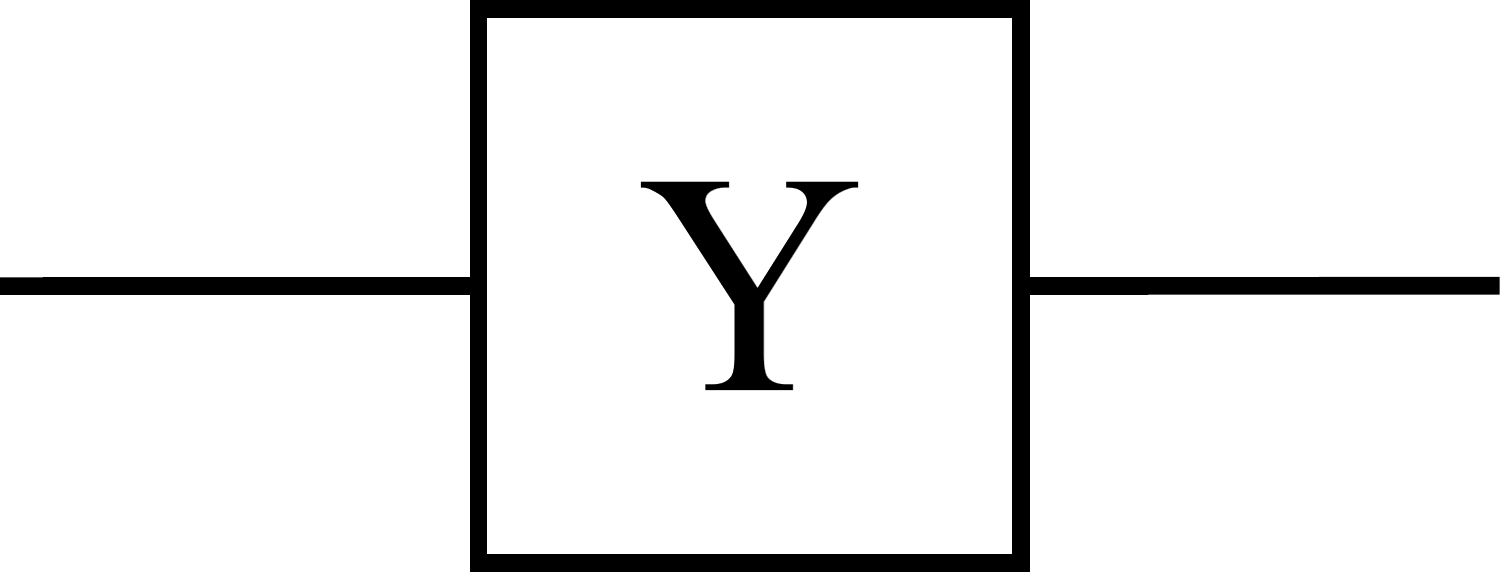}} \\
Pauli Z ($Z$)    &    \parbox[c]{7cm}{\centering\includegraphics[scale=0.075]{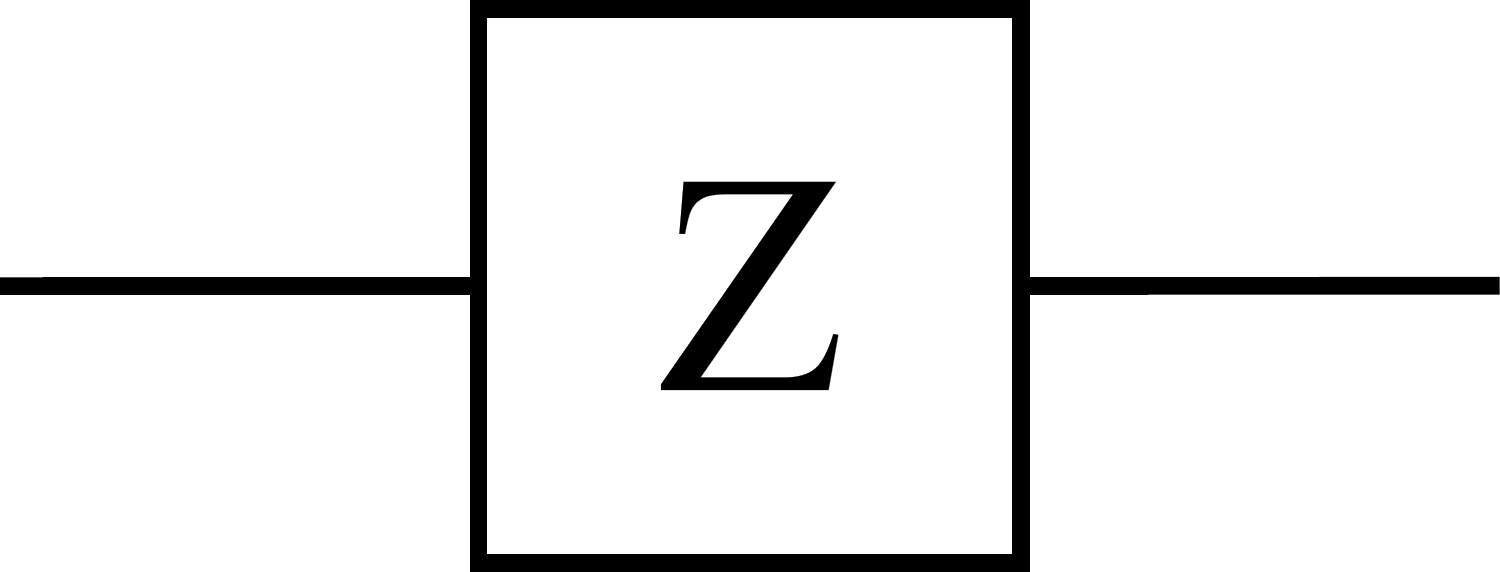}} \\
Hadamard ($H$)   &    \parbox[c]{7cm}{\centering\includegraphics[scale=0.075]{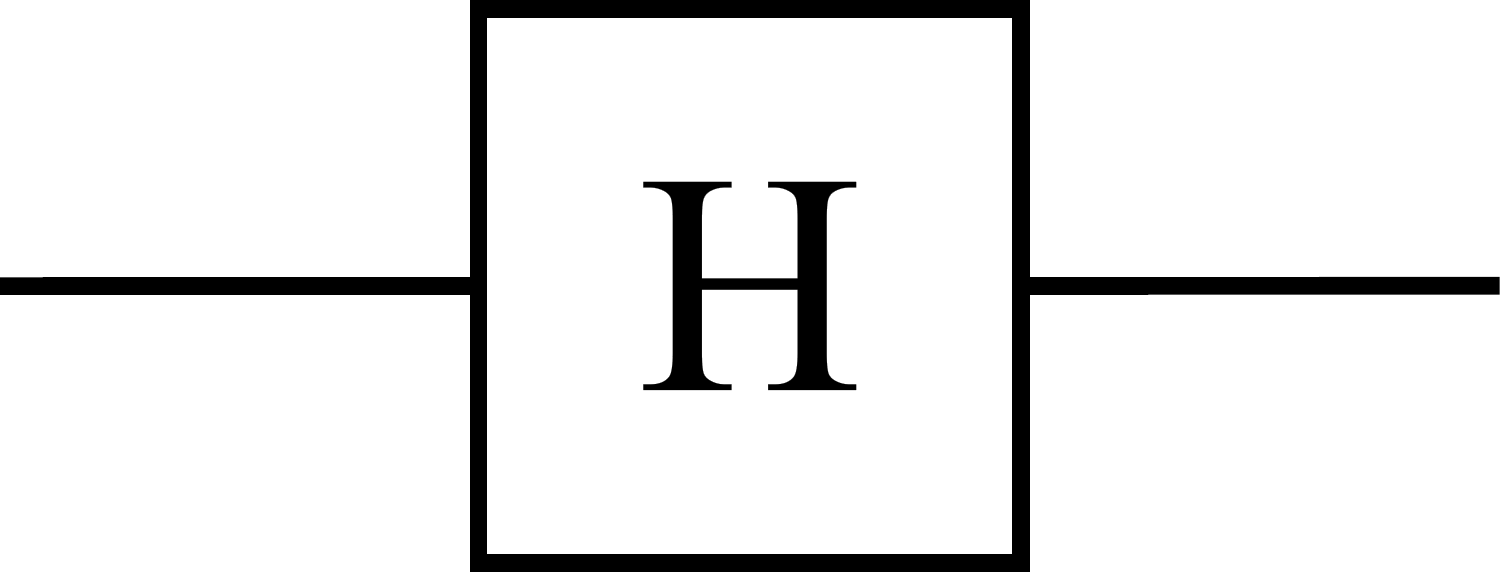}} \\
Phase ($S$, $P$)   &    \parbox[c]{7cm}{\centering\includegraphics[scale=0.075]{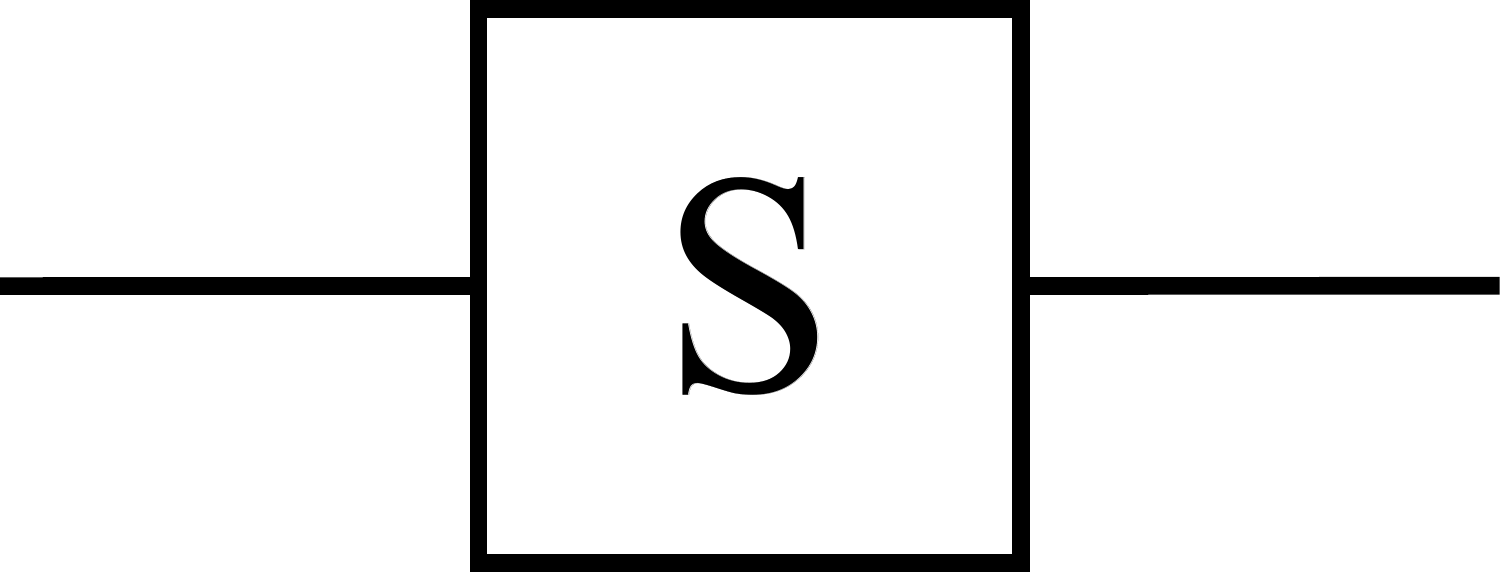}}\\
$\pi/8$ ($T$)    &    \parbox[c]{7cm}{\centering\includegraphics[scale=0.075]{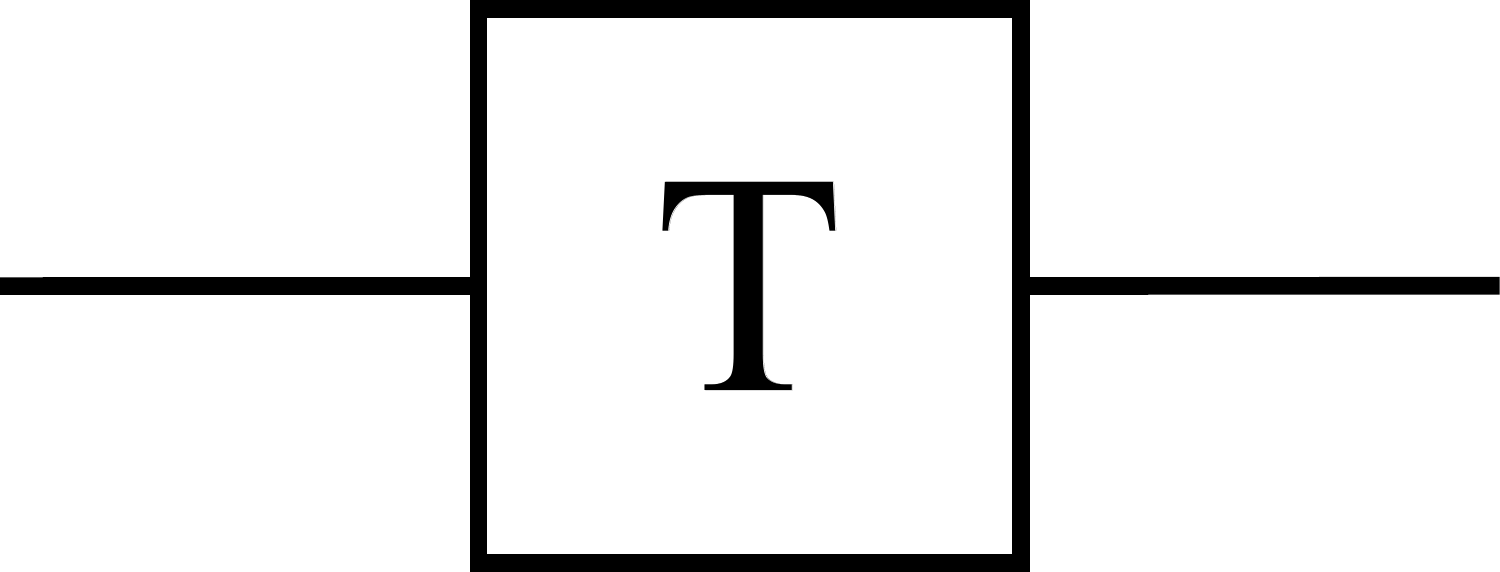}}\\
Controlled Not (CNOT) & \parbox[c]{7cm}{\centering\includegraphics[scale=0.075]{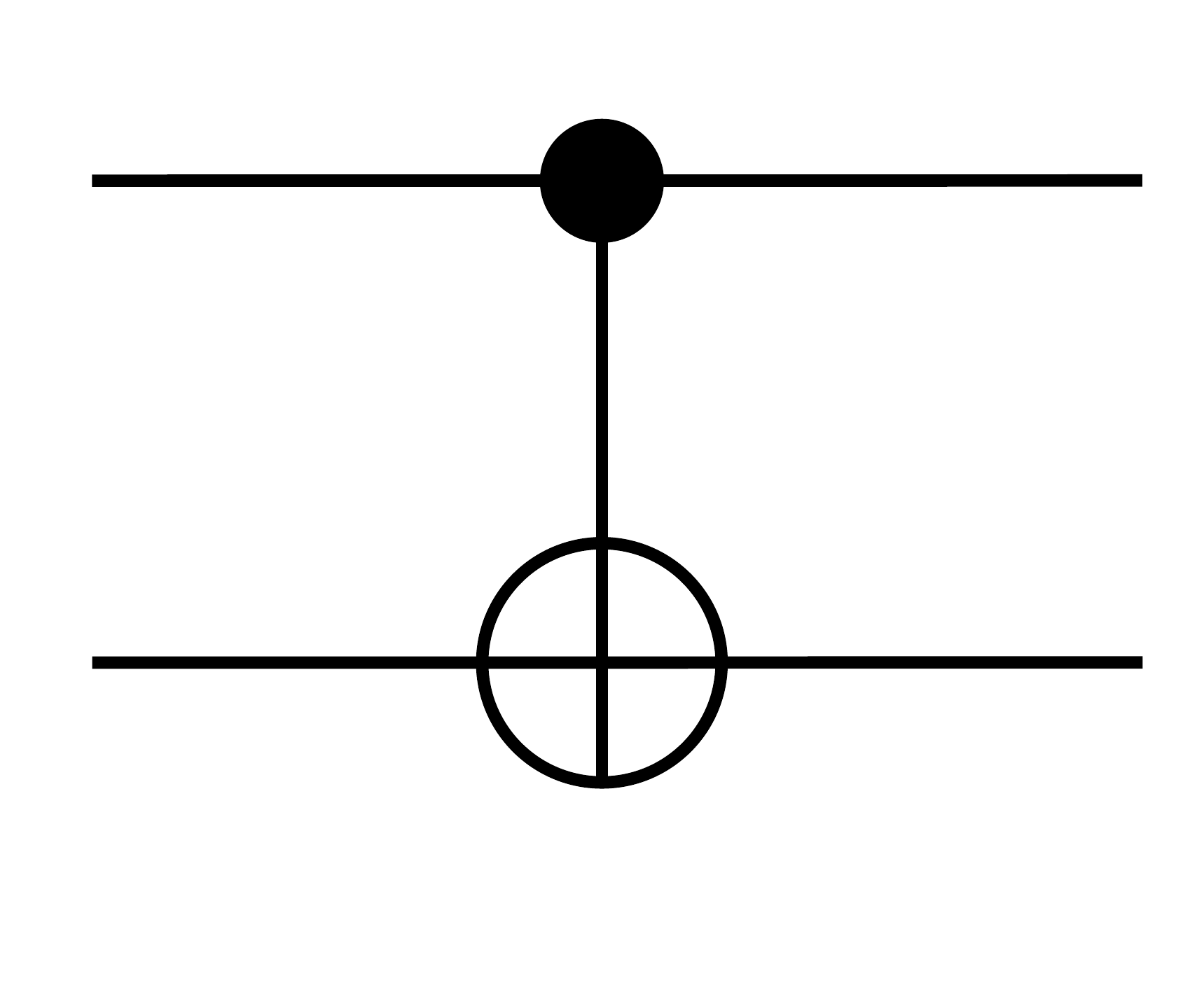}} \\
Controlled Z (CZ)         & \parbox[c]{7cm}{\centering\includegraphics[scale=0.075]{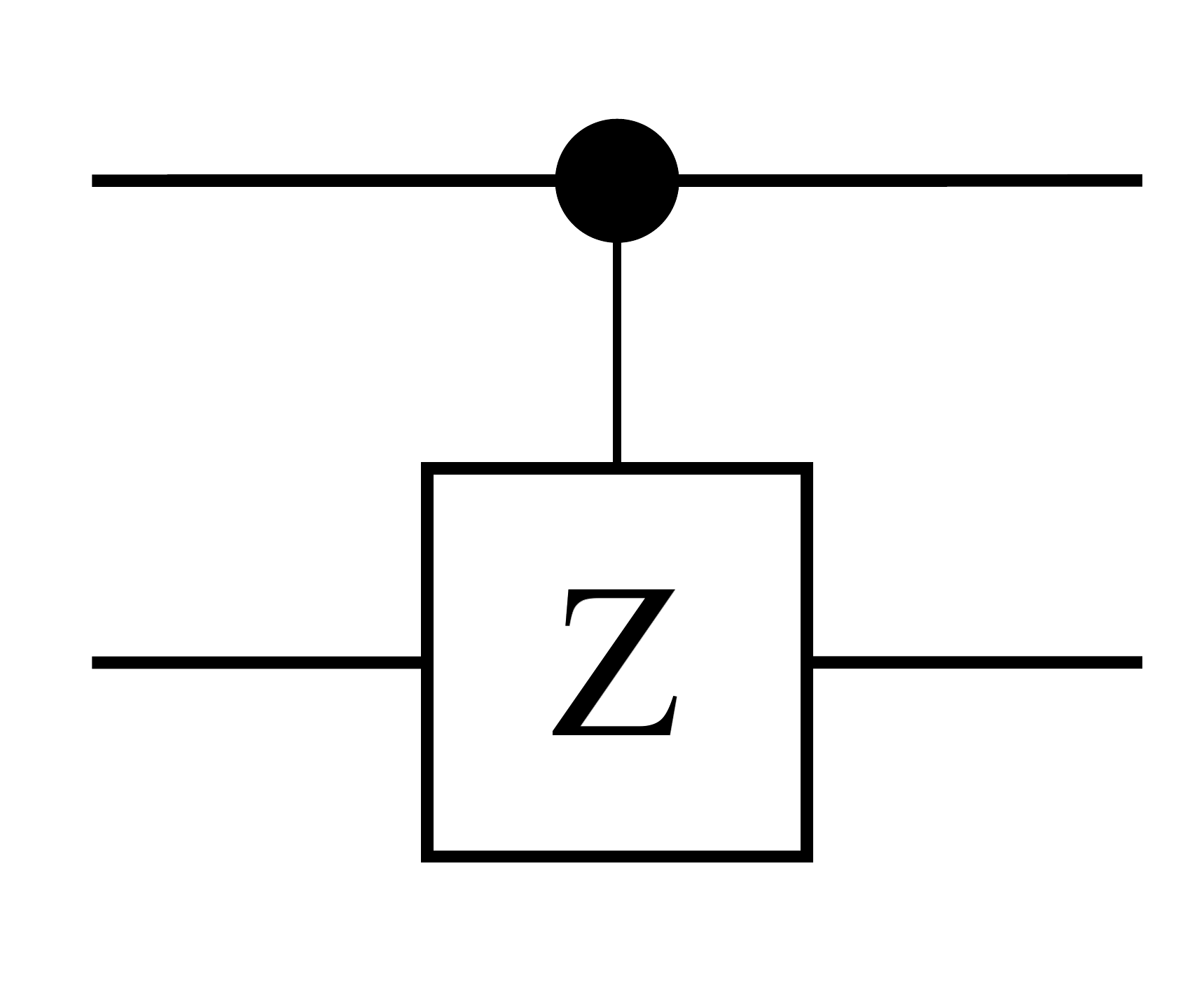} \includegraphics[scale=0.075]{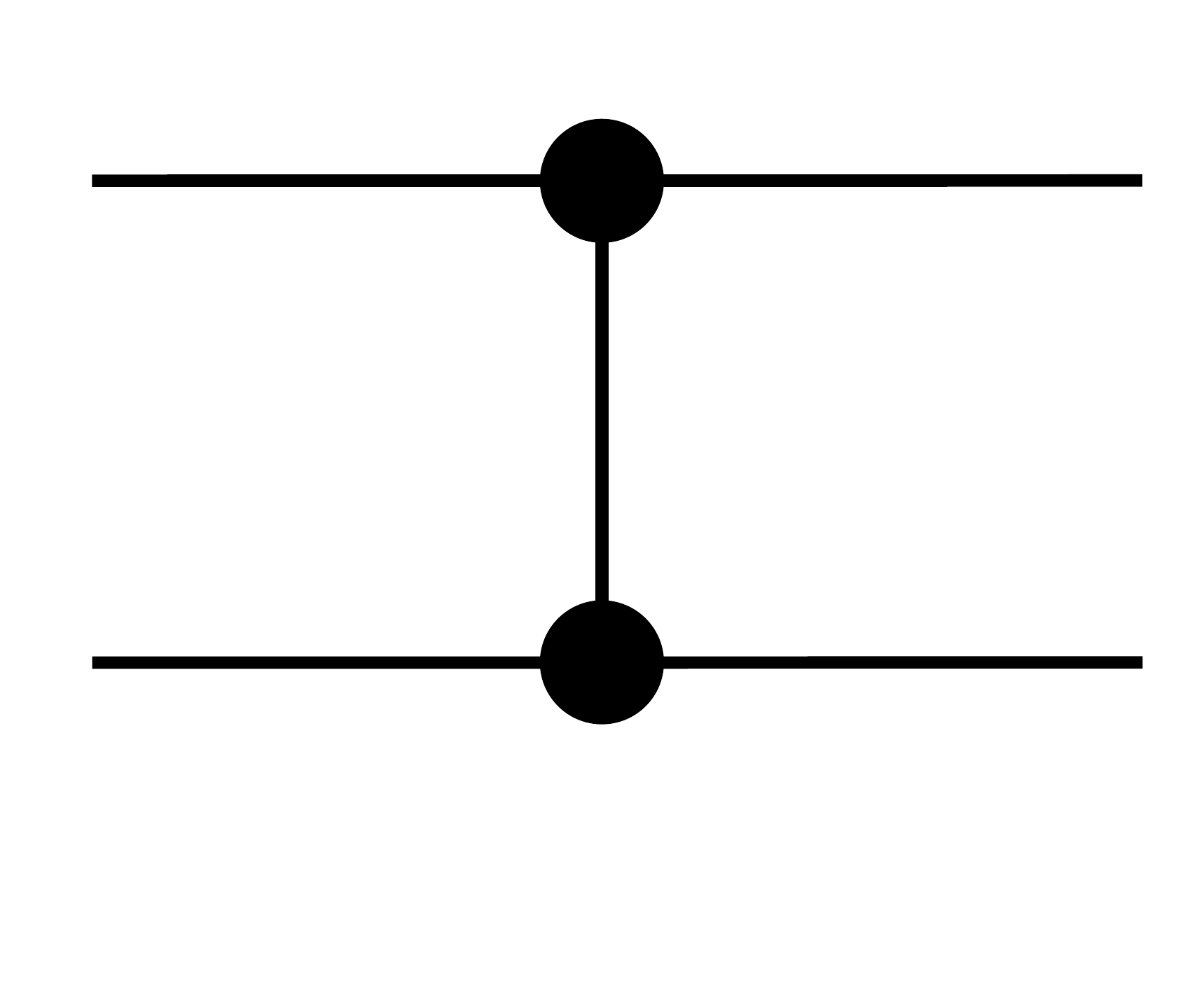}} \\
SWAP                      & \parbox[c]{7cm}{\centering\includegraphics[scale=0.075]{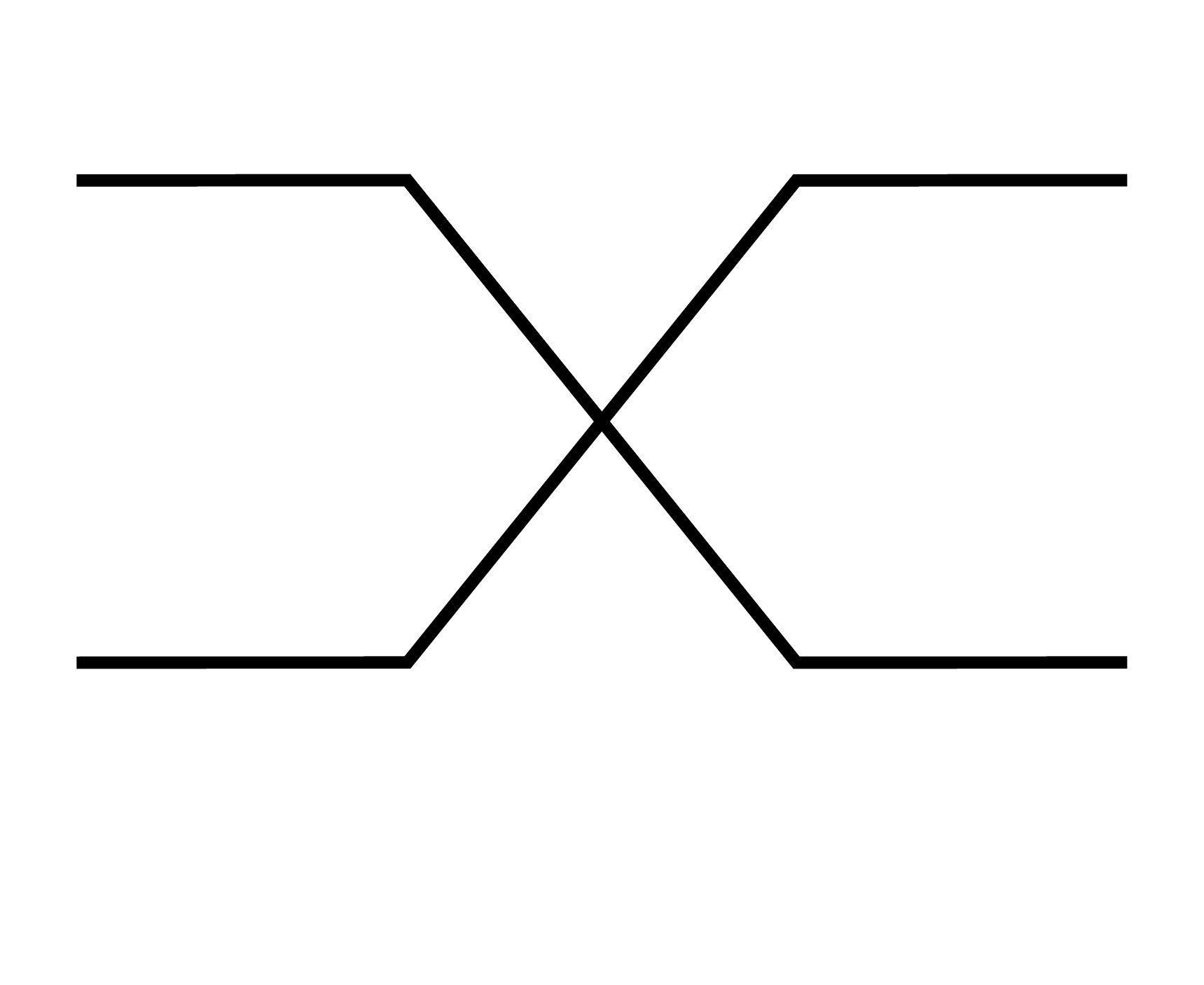} \includegraphics[scale=0.075]{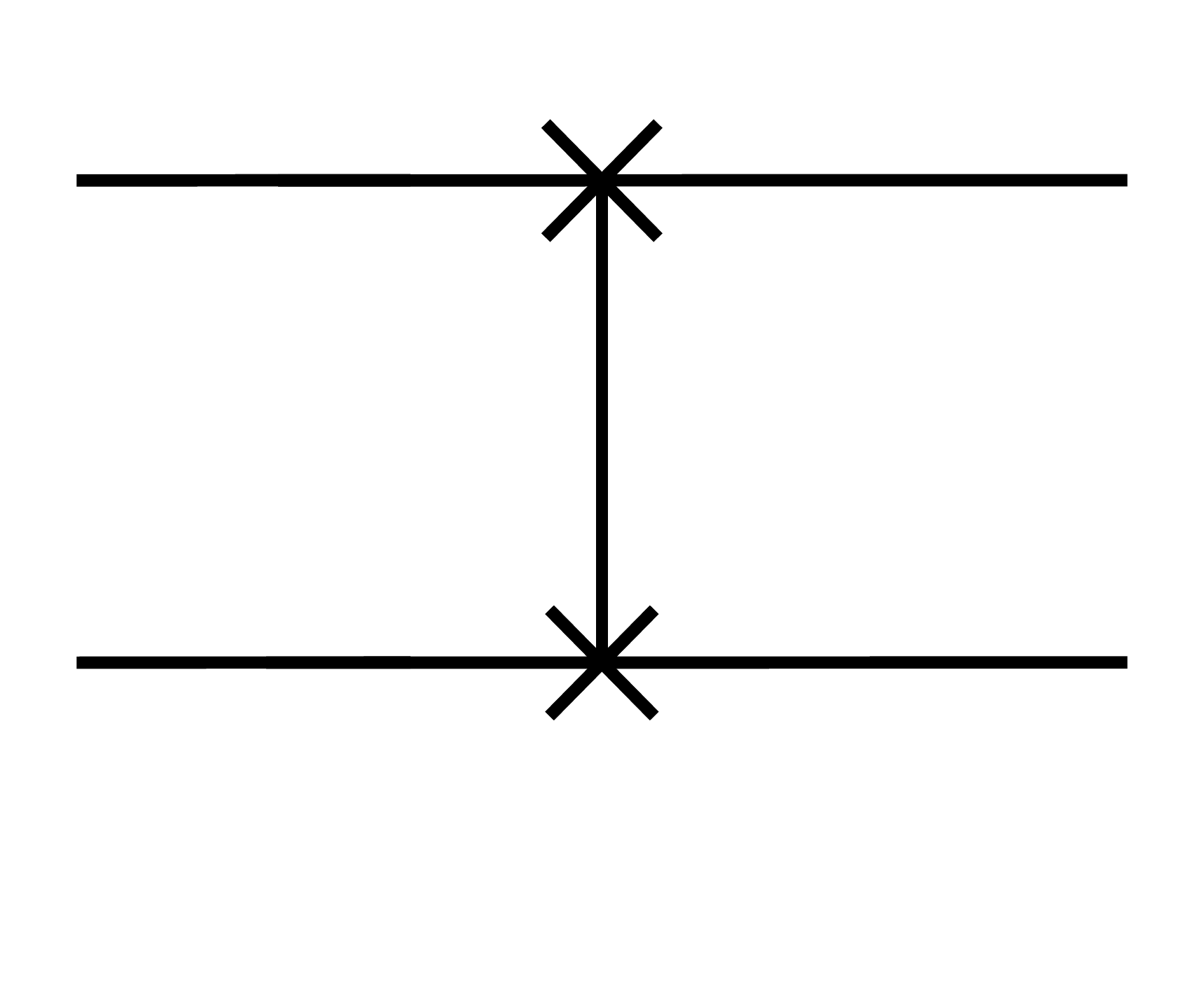}}         
\end{tabular}
\egroup
\caption{Quantum gates and corresponding circuit representations used in our evaluation of the capabilities of \chatgpt{} regarding quantum computing. These gates are considered as \textit{common} by Wikipedia \citep{WikipediaQuantumGate}.}
\label{t:circuits}
\end{table}

\end{document}